\documentclass[twoside,twocolumn,english]{revtex4-1}

\usepackage[T1]{fontenc}
\usepackage[latin9]{inputenc}
\usepackage{amsmath}
\usepackage{amssymb}
\usepackage{graphicx}

\makeatletter
\usepackage[spanish]{babel}

\makeatother

\usepackage{babel}
\begin{document}

\title{Molecular emission near metal interfaces: the polaritonic regime}

\author{Joel Yuen-Zhou$^{1}$, Semion K. Saikin$^{2}$, Vinod M. Menon$^{3}$}

\affiliation{$^{1}$Department of Chemistry and Biochemistry, University of California
San Diego, La Jolla, CA, USA.}
\email{joelyuen@ucsd.edu}

\affiliation{$^{2}$Department of Chemistry and Chemical Biology, Harvard University,
Cambridge, MA, USA.}

\affiliation{$^{2}$Institute of Physics, Kazan Federal University, Kazan, Russian
Federation.}

\affiliation{$^{4}$Department of Physics, Graduate Center and City College of
New York, City University of New York, New York, New York, USA. }
\begin{abstract}
The strong coupling of a dense layer of molecular excitons with surface-plasmon
modes in a metal gives rise to polaritons (hybrid light-matter states)
called plexcitons. Surface plasmons cannot directly emit into (or
be excited by) free-space photons due to the fact that energy and
momentum conservation cannot be simultaneously satisfied in photoluminescence.
Most plexcitons are also formally non-emissive, even though they can
radiate via molecules upon localization due to disorder and decoherence.
However, a fraction of them are bright even in the presence of such
deleterious processes. In this letter, we theoretically discuss the
superradiant emission properties of these bright plexcitons, which
belong to the upper energy branch and reveal huge photoluminescence
enhancements compared to bare excitons. Our study generalizes the
well-known problem of molecular emission next to a metal interface
to collective molecular states and provides new design principles
for the control of photophysical properties of molecular aggregates
using polaritonic strategies.
\end{abstract}
\maketitle
The study of molecular photoluminescence (PL) next to a metal-dielectric
interface dates back to a classic experiment reported by Drexhage,
Kuhn, and coworkers almost fifty years ago \cite{drexhage1968variation,drexhage1970influence}.
By controlling the distance between a molecular emitter and a metal-mirror
film, the aforementioned authors showed that the observed PL rate
oscillated and then monotonically increased for short distances. These
oscillations were attributed to interferences between the free-space
and reflected light from the mirror, and the monotonic decay was associated
to irreversible energy transfer to surface plasmons (SPs) in the metal.
These observations were later fully elucidated in a theory provided
by Chance, Prock, and Silbey (CPS) \cite{chance1974lifetime,chance1978molecular},
by adapting the results of an even older problem of antenna radiation
next to the surface of the Earth, whose mathematical solution was
offered by Sommerfeld as early as 1909 \cite{sommerfeld1909ausbreitung}.
This problem has been revisited countless times to understand molecular
energy transfer processes in condensed phases.

\begin{figure}
\centering{}\includegraphics[scale=0.3]{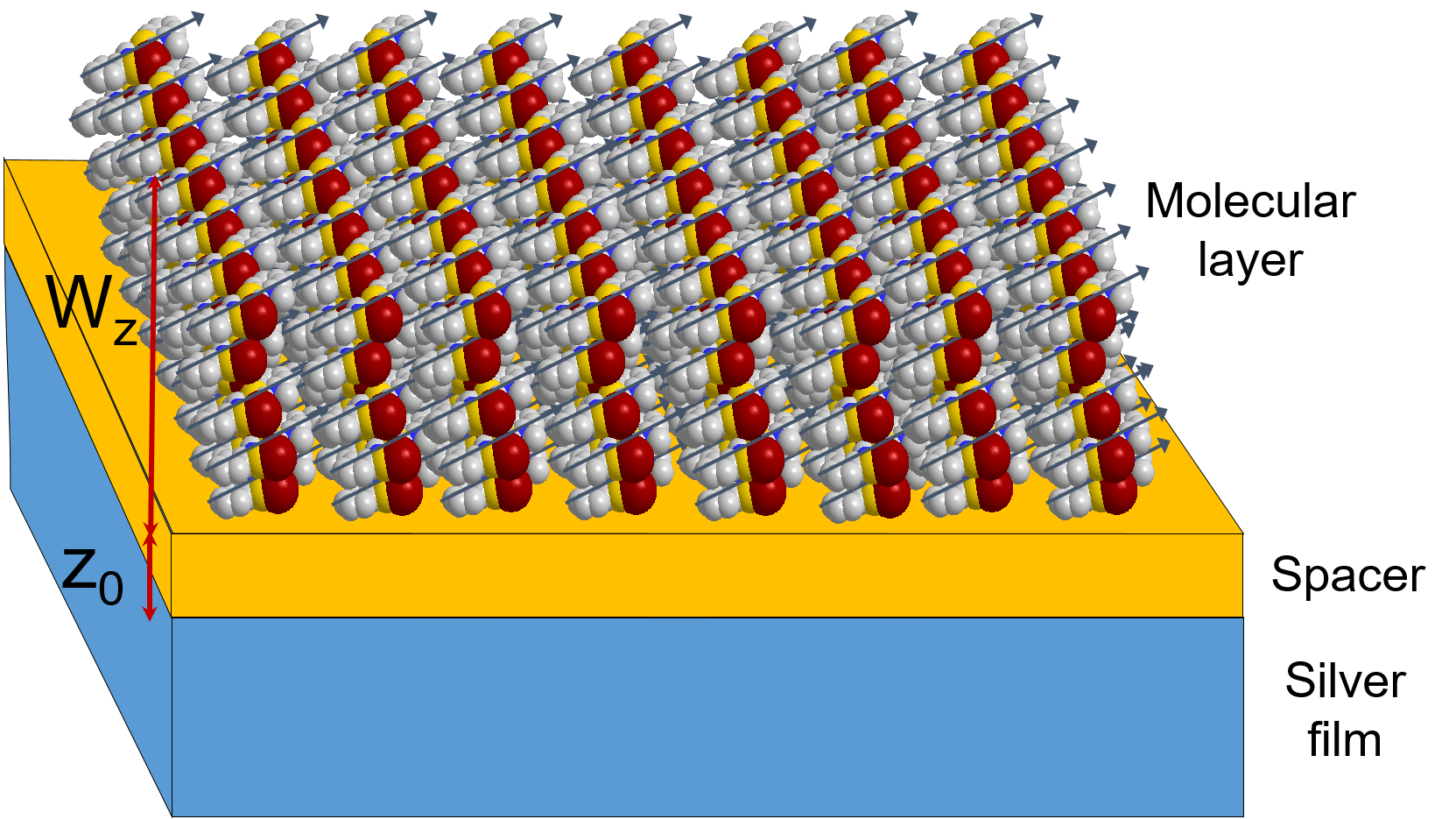}\caption{\emph{Plexciton setup.} A molecular layer of thickness $W_{z}$ sits
on top of a thin dielectric spacer of thickness $z_{0}$, which in
turn, is placed on a metal film. Strong coupling between excitons
in the molecular layer and surface-plasmons (SPs) in the metal film
imply energy exchange between the excitons and SPs is much faster
than their respective decay processes, giving rise to polaritonic
excitations called plexcitons.\label{fig:Plexciton-setup}}
\end{figure}

In this letter, we study a variation of the aforementioned problem
which offers new phenomenology that, as far as we are aware, has not
been reported before. We study PL originating from (delocalized) molecular
excited states (excitons) when they are strongly coupled to a SP metal
film (see Fig. \ref{fig:Plexciton-setup}). We consider a layer of
$N_{x}N_{y}N_{z}$ molecules (of thickness $W_{z}$) placed on top
of a thin dielectric spacer (of thickness $z_{0}$), which is on a
thick metal film. When the molecular layer is sufficiently dense,
the energy transfer between the exciton and SP modes is reversible
and faster than each of the decay rates \cite{gonzalez_tudela,torma2015strong}.
The resulting eigenmodes are no longer purely plasmonic or excitonic,
but rather, polaritonic, being coherent superpositions of such modes.
Polaritons arising from SPs and excitons are termed \emph{plexcitons}
\cite{PhysRevLett.93.036404,halas,ozel2013observation,kochuveedu2014surface,yuen2016plexciton}.
As with any polaritonic system, the transmission spectrum of the plexciton
system at the wavevector giving rise to resonance between the uncoupled
exciton and SP bands shows a Rabi splitting (anticrossing) between
two new bands, named \emph{upper} and \emph{lower} plexcitons (UPs,
LPs), according to their energy ordering \cite{torma2015strong}.
Molecular polaritons have recently been the subject of intense investigation,
as they offer new ways for coherent control of molecular processes
\cite{brumer} such as changes on reaction rates and thermodynamics
by ``imprinting'' electromagnetic coherence directly onto the molecular
states \cite{ebbesen2016hybrid,herrera2016cavity,galego2016suppressing,Flick2017,baranov2017novel,sukharev2017optics,ribeiro2018polariton}.
They also provide new platforms to induce remote energy transfer \cite{Zhong2017,du2018theory,saez2018organic}
and to recreate exotic topological \cite{yuen2016plexciton,gao2018continuous}
and many-body phenomena at room temperature in tabletop experiments
\cite{bittner_silva,daskalakis2014nonlinear,zaster2016quantum}. In
this study, we highlight superradiant properties \cite{dicke1954coherence,spano1989superradiance}
that plexcitons exhibit which are not encountered in the weak coupling
regime of a bare molecule next to a metal surface. These properties
can potentially be harnessed for light harvesting and energy routing
purposes in molecular materials and device applications.

\begin{figure}
\centering{}\includegraphics[scale=0.35]{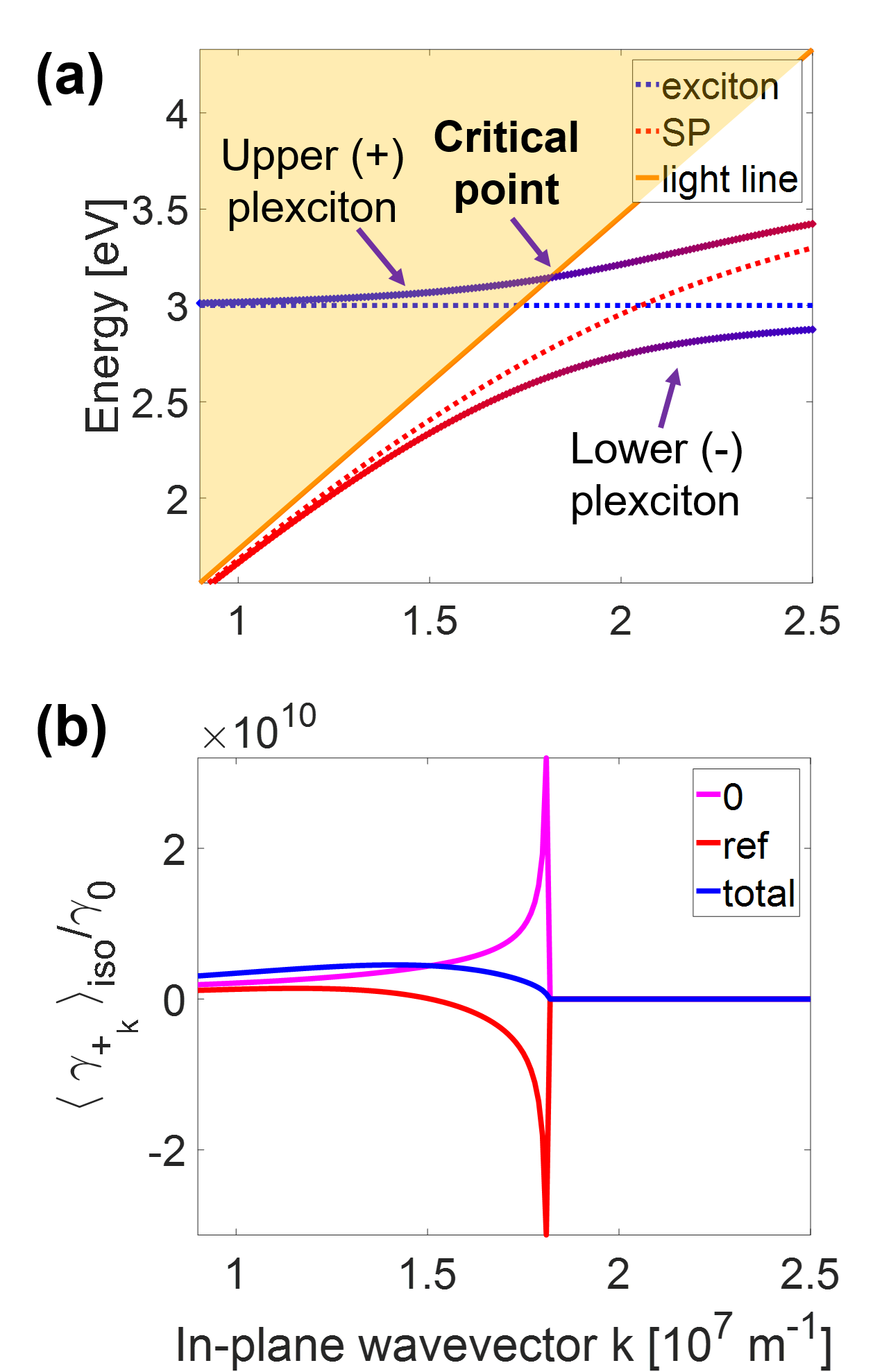}\caption{(a) \emph{Dispersion plot of excitations for an isotropic molecular
layer. }When the (flat) exciton (dashed blue) and the SP (dashed red)
bands couple, they give rise to upper and lower plexciton (UP, LP).
The UP $|+_{\boldsymbol{k}}\rangle$ interpolates between exciton
and SP between $|\boldsymbol{k}|\approx0$ and $|\boldsymbol{k}|\to\infty$;
the converse is true for the LP ($|-_{\boldsymbol{k}}\rangle$). Strong
hybridization occurs in the anticrossing region. The yellow-shaded
region contains states that emit photons, while the right region contains
states that cannot due to mismatch in energy and in-plane wavevector
with respect to free-space photons. The critical point at $\boldsymbol{k}=\boldsymbol{k}^{*}$separates
bright and non-emissive UPs. (b) \emph{PL rates for the UP band $\langle\gamma_{+_{\boldsymbol{k}}}\rangle_{iso}=\sum_{i}\langle\gamma_{+_{\boldsymbol{k}},i}\rangle_{iso}$
normalized to the single-molecule rate $\gamma_{SM}$}. Contributions
to $\langle\gamma_{+_{\boldsymbol{k}}}\rangle_{iso}$ correspond to
direct ($i=0$), and interference terms between free-space and reflected
waves ($i=ref$). The divergences in $\langle\gamma_{+_{\boldsymbol{k}^{*}},i}\rangle_{iso}$
are van Hove singularities. \label{fig:dispersion}}
\end{figure}

We first briefly lay out a quantum-mechanical formalism to describe
the plexciton system (see details in Supporting Information, SI-\ref{sec:SP-modes:-properties},
\ref{sec:Derivation-of-plexciton}). Writing the Hamiltonian as $H=\sum_{\boldsymbol{k}}H_{\boldsymbol{k}}$,
where the sum runs over a discrete set of in-plane exciton wavevectors
$\boldsymbol{k}$, we have \cite{gonzalez_tudela,torma2015strong}

\begin{align}
H_{\boldsymbol{k}} & =\hbar\omega_{\boldsymbol{k}}^{SP}a_{\boldsymbol{k}}^{\dagger}a_{\boldsymbol{k}}+\hbar\omega_{e}\sigma_{\boldsymbol{k}}^{\dagger}\sigma_{\boldsymbol{k}}\nonumber \\
 & +(\mathcal{J}_{\boldsymbol{k}}\sigma_{\boldsymbol{k}}^{\dagger}a_{\boldsymbol{k}}+\mbox{h.c.})+H_{dark,\boldsymbol{k}}+H_{umklapp,\boldsymbol{k}}.\label{eq:Hk}
\end{align}
Here, $\hbar\omega_{\boldsymbol{k}}^{SP}$ and $a_{\boldsymbol{k}}^{\dagger}(a_{\boldsymbol{k}})$
{[} $\hbar\omega_{e}$ and $\sigma_{\boldsymbol{k}}^{\dagger}(\sigma_{\boldsymbol{k}})${]}
are the energy and creation (annhilation) operator of a $\boldsymbol{k}$th
SP {[}exciton{]}; $\mathcal{J}_{\boldsymbol{k}}$ is the collective
SP-exciton coupling. $H_{dark,\boldsymbol{k}}$ describes dark excitons
at the bare exciton energy $\hbar\omega_{e}$, which do not couple
directly to SPs \cite{ribeiro2018polariton}; we disregard $H_{umklapp,\boldsymbol{k}}$
as a negligible off-resonant contribution due to coupling with high-wavevector
SP modes. Ignoring also $H_{dark,\boldsymbol{k}}$ for the time-being
(we will discuss it at the end of the letter), $H_{\boldsymbol{k}}$
in Eq. (\ref{eq:Hk}) becomes a two-level system for every $\boldsymbol{k}$,
and can be diagonalized to yield two plexciton (polariton) states
of the form $|y_{\boldsymbol{k}}\rangle=\Big[\zeta_{y_{\boldsymbol{k}}}^{(SP)}a_{\boldsymbol{k}}^{\dagger}+\zeta_{y_{\boldsymbol{k}}}^{(exc)}\sigma_{\boldsymbol{k}}^{\dagger}\Big]|\mbox{vac}\rangle$
with eigenenergies $\hbar\omega_{y_{\boldsymbol{k}}}$, where $y=\pm$
are labels for the UP (LP), and $|\mbox{vac}\rangle=|g;0_{SP};0_{UHP}\rangle$
is the tensor product of the ground state for the molecular degrees
of freedom ($|g\rangle$) and the vacuum for the SP modes ($|0_{SP}\rangle$);
see Fig. \ref{fig:dispersion}a.

Fig. \ref{fig:dispersion}a shows a dispersion energy plot for exciton,
SP, and resulting plexciton bands assuming $\omega_{e}=3\,\text{eV}$,
$|\boldsymbol{\mu}_{eg}|=10\,\text{Debye}$, and relative permittivities
$\epsilon_{d}=1.3$ and $\epsilon_{m}(\omega)=3.7-\frac{(8.6\,\text{eV})^{2}}{\omega^{2}}$
for the molecular (dielectric) layer and metal film, respectively.
We used $\rho=10^{9}\,\text{molecules}/\mu\text{m}^{-3}$, $z_{0}=1\,\text{nm}$,
and $W_{z}=200\,\text{nm}$, to account for representative parameters
in the literature \cite{gonzalez_tudela}. We took isotropic averages
of $\mathcal{J}_{\boldsymbol{k}}$ to describe a molecular layer with
orientational disorder. Conservation of energy and in-plane momentum
constrains the allowed PL processes: an excitation of frequency $\omega$
and in-plane wavevector $\boldsymbol{k}$ must output a propagating
photon with the same properties. This means that, for PL to occur,
there needs to be a real-valued $k_{zd}$ satisfying
\begin{align}
\omega & =\frac{c}{\sqrt{\epsilon_{d}}}\sqrt{|\boldsymbol{k}|^{2}+|k_{zd}|^{2}}.\label{eq:conservation}
\end{align}
It is well known this is not a possibility for SP modes \cite{maier},
whose dispersion is to the right of the ``light-line,'' $\omega_{\boldsymbol{k}}^{SP}<\omega_{\boldsymbol{k}}^{ll}\equiv\frac{c}{\sqrt{\epsilon_{d}}}|\boldsymbol{k}|$,
and thus, must be probed by coupling to gratings \cite{novotny} or
nanoparticles \cite{aravind1982use}, for example. The same holds
formally true for LPs ($\omega_{-_{\boldsymbol{k}}}<\omega_{\boldsymbol{k}}^{ll}$)
and for a subset of UPs; to distinguish them from the dark exciton
states (eigenstates of $H_{dark,\boldsymbol{k}}$), we shall call
them non-emissive states. Importantly, there is another subset of
UPs that are to the left of the light-line, thus being formally bright
and featuring PL (see Fig. \ref{fig:dispersion}a). We then ask: what
is the rate of PL from these UPs, and how does the bright-to-non-emissive
transition occur as a function of $\boldsymbol{k}$? To answer these
questions, we compute the PL rate of a state $|y_{\boldsymbol{k}}\rangle$
into the free-space radiative modes of the upper-half-plane (UHP)
of the plexciton setup. The corresponding Wigner-Weiskopf expression
(Fermi golden rule) is \cite{schatz1993quantum,Scully1997}

\begin{align}
\gamma_{y_{\boldsymbol{k}}} & =\frac{2\pi}{\hbar}\sum_{\boldsymbol{K},\chi}\left|\langle\mbox{vac};(\boldsymbol{K},\chi)_{UHP}|H_{int}|y_{\boldsymbol{k}};0_{UHP}\rangle\right|^{2}\delta(\hbar\omega_{y_{\boldsymbol{k}}}-\hbar\omega_{\boldsymbol{K}}^{UHP}),\label{eq:gamma_sk}
\end{align}
where $H_{int}=-\sum_{\boldsymbol{n}s}\hat{\boldsymbol{\mu}}_{\boldsymbol{n}s}\cdot\hat{\boldsymbol{\mathcal{E}}}_{UHP}(\boldsymbol{r}_{\boldsymbol{n}s})$
contains the interaction between each of the molecules in the slab
and the electric field at the UHP. In Eq. (\ref{eq:gamma_sk}), $|0_{UHP}\rangle$
denotes the UHP photonic vacuum state. The matrix element couples
an initial state with a plexciton and no UHP photons $|y_{\boldsymbol{k}};0_{UHP}\rangle$
with a final state featuring no plexcitons and an UHP photon with
energy $\hbar\omega_{\boldsymbol{K}}^{UHP}=\frac{\hbar c|\boldsymbol{K}|}{\sqrt{\epsilon_{d}}}$,
$|\mbox{vac};(\boldsymbol{K},\chi)_{UHP}\rangle=b_{\boldsymbol{K},\chi}^{\dagger}|\mbox{vac};0_{UHP}\rangle$,
where $b_{\boldsymbol{K},\chi}^{\dagger}(b_{\boldsymbol{K},\chi})$
is the creation (annhilation) operator of photons at the UHP mode
with (three-dimensional) wavevector $\boldsymbol{K}=(K_{x},K_{y},K_{zd})$
and polarization index $\chi=s,p$, which satisfies the commutation
relation $[b_{\boldsymbol{K},\chi},b_{\boldsymbol{K'},\chi'}^{\dagger}]=\delta_{\boldsymbol{K},\boldsymbol{K'}}\delta_{\chi,\chi'}$.
The electric field at the UHP is given by a collection of radiative
modes. Using the modal representation given by Arnoldus and George
\cite{arnoldus1988spontaneous},

\begin{align}
\hat{\boldsymbol{\mathcal{E}}}_{UHP}(\boldsymbol{r}) & =\sum_{\boldsymbol{K},\chi}\Theta(-\boldsymbol{K}\cdot\hat{\boldsymbol{z}})\left\{ \frac{(b_{\boldsymbol{K},\chi}+b_{\tilde{\boldsymbol{K}},\chi})}{\sqrt{1+|r_{\boldsymbol{K}\chi}|^{2}}}\sqrt{\frac{\hbar\omega_{\boldsymbol{K}}^{UHP}}{2\epsilon_{0}\epsilon_{d}V}}\right.\nonumber \\
 & \left.\times\Bigg[\boldsymbol{e}_{\boldsymbol{K},\chi}e^{i\boldsymbol{K}\cdot\boldsymbol{r}}+r_{\boldsymbol{K},\chi}e^{i\tilde{\boldsymbol{K}}\cdot\boldsymbol{r}}\boldsymbol{e}_{\tilde{\boldsymbol{K}},\chi}\Bigg]+\mbox{h.c.}\right\} .\label{eq:E_UHP}
\end{align}
The radiative modes can have $s$ or $p$ polarization, $\boldsymbol{e}_{\boldsymbol{K},s}=\frac{\boldsymbol{K}\times\hat{\boldsymbol{z}}}{|\boldsymbol{K}\times\hat{\boldsymbol{z}}|}=\frac{K_{y}\hat{\boldsymbol{x}}-K_{x}\hat{\boldsymbol{y}}}{\sqrt{K_{x}^{2}+K_{y}^{2}}}$
and $\boldsymbol{e}_{\boldsymbol{K},p}=\frac{\boldsymbol{e}_{\boldsymbol{K}s}\times\boldsymbol{K}}{|\boldsymbol{e}_{\boldsymbol{K}s}\times\boldsymbol{K}|}=\frac{-K_{zd}(K_{x}\hat{\boldsymbol{x}}+K_{y}\hat{\boldsymbol{y}})}{|\boldsymbol{K}|\sqrt{K_{x}^{2}+K_{y}^{2}}}+\frac{\sqrt{K_{x}^{2}+K_{y}^{2}}\hat{\boldsymbol{z}}}{|\boldsymbol{K}|}$.
$V$ is the quantization volume of the UHP, $\Theta$ is the Heavyside
step function, and $\tilde{\boldsymbol{K}}=\boldsymbol{K}-2\boldsymbol{K}\cdot\hat{\boldsymbol{z}}\hat{\boldsymbol{z}}$
is the reflected wavevector for an incident wave into the metal with
$K_{zd}<0$ and Fresnel coefficients $r_{\boldsymbol{K},s}=\frac{K_{zd}-K_{zm}}{K_{zd}+K_{zm}}$
and $r_{\boldsymbol{K},p}=\frac{\epsilon_{m}K_{zd}-\epsilon_{d}K_{zm}}{\epsilon_{m}K_{zd}+\epsilon_{d}K_{zm}}$
\cite{novotny}, where $K_{zm}(\boldsymbol{K},\omega)=-\sqrt{\epsilon_{m}(\omega)\frac{\omega^{2}}{c^{2}}-|\boldsymbol{K}_{\perp}|^{2}}$.
Note that although $K_{zd}\in\Re$, $K_{zm}\in\Im$ given that the
UP frequency is typically below the asymptotic $k\to\infty$ SP frequency,
$\omega_{\boldsymbol{K}}^{UHP}<\frac{\omega_{P}}{\sqrt{\epsilon_{d}+\epsilon_{\infty}}}$.
This in turn yields $|r_{\boldsymbol{K}\chi}|^{2}=1$,\emph{ i.e.},
lossless metals are perfectly reflective \cite{arnoldus1988spontaneous}.

Eq. (\ref{eq:E_UHP}) implies the rate $\gamma_{y_{\boldsymbol{k}}}$
calculated through Eq. (\ref{eq:gamma_sk}) can be partitioned into
two contributions, $\gamma_{y_{\boldsymbol{k}}}=\gamma_{y_{\boldsymbol{k}},0}+\gamma_{y_{\boldsymbol{k}},ref}$,
corresponding to the direct term associated with the free-space and
reflected electric fields independently, and the interference term
between them, as done with single molecule calculations \cite{chance1978molecular,Ford1984,barnes1998fluorescence}.
We present results normalized to $\gamma_{SM,0}=\frac{\sqrt{\epsilon_{d}}}{3\pi\epsilon_{0}\hbar}\left(\frac{\omega_{e}}{c}\right)^{3}|\boldsymbol{\mu}_{eg}|^{2},$
the single-molecule PL rate in free-space, which attains a value of
$0.5\,\text{ns}^{-1}$ for the present parameters. Recall that the
single-molecule PL rate scales as $\omega_{e}^{3}$ because it can
radiate across all solid angles (the number of accesible radiative
modes grows as $\omega_{e}^{2}$) and it is proportional to the intensity
of the electric field ($\propto\omega_{e}$) \cite{schatz1993quantum,Scully1997}.
The evaluation of Eq. (\ref{eq:gamma_sk}) is presented in SI-\ref{sec:Evaluation-of-plexciton},
where we also show in great detail an alternative route to obtain
the same results using the dyadic Green's function formalism \cite{kong1975theory,Tai1994,novotny}.
Here, we limit ourselves to discuss the isotropically distributed
molecular layer, which already captures the essential features of
the general solution; denoting isotropic averages by $\langle\cdot\rangle_{iso}$,
we obtain (see Fig. \ref{fig:dispersion}b),
\begin{align}
\frac{\langle\gamma_{y_{\boldsymbol{k}},i}\rangle_{iso}}{\gamma_{SM,0}} & =\left(\frac{2\pi c}{\sqrt{\epsilon_{1}}}\right)|\zeta_{y_{\boldsymbol{k}}}^{(exc)}|^{2}\Theta\left(\omega_{y_{\boldsymbol{k}}}-\frac{c}{\sqrt{\epsilon_{d}}}|\boldsymbol{k}|\right)\rho\label{eq:ratio}\\
 & \times\left(\frac{\omega_{y_{\boldsymbol{k}}}^{2}\frac{1}{k_{zd}}}{\omega_{e}^{3}}\right)\frac{f_{i}}{\int_{z_{0}}^{z_{f}}dz|J_{\boldsymbol{k}}(z)|^{2}}\nonumber
\end{align}
where\begin{subequations}\label{eq:f}

\begin{align}
f_{0} & =\left|\int_{z_{0}}^{z_{f}}dzJ_{\boldsymbol{k}}(z)e^{-ik_{zd}z}\right|^{2},\label{eq:f0}\\
f_{ref} & =\frac{1}{2}\Re\Bigg\{\left(r_{\boldsymbol{k}-k_{zd}\hat{\boldsymbol{z}},s}-r_{\boldsymbol{k}-k_{zd}\hat{\boldsymbol{z}},p}\frac{k_{dz}^{2}-|\boldsymbol{k}|^{2}}{k_{dz}^{2}+|\boldsymbol{k}|^{2}}\right)\nonumber \\
 & \times\Big[\int_{z_{0}}^{z_{f}}dzJ_{\boldsymbol{k}}(z)e^{ik_{zd}z}\Big]^{2}\Bigg\}.\label{eq:fref}
\end{align}
\end{subequations}Eqs. (\ref{eq:ratio}) and (\ref{eq:f}) are the
main results of this letter. To capture the essential physics behind
the PL trends, we have so far considered a lossless metal. However,
as we demonstrate in SI-\ref{sec:Effects-of-metal}, the formalism
of macroscopic quantum electrodynamics (M-QED) \cite{dung2000spontaneous,vogel2006quantum,scheel2008macroscopic}
validates these expressions even when the metal is lossy, where |$r_{\boldsymbol{k}-k_{zd}\hat{\boldsymbol{z}},\chi}|<1$.
Let us provide a physical interpretation of these results. The PL
rate for a plexciton state is proportional to its exciton population
$|\zeta_{y_{\boldsymbol{k}}}^{(exc)}|^{2}$, given the SP population
does not participate in the process. The factor of molecular density
$\rho$ signals the onset of superradiance, where the PL rate scales
as the number of molecules in the coherently delocalized state, analogously
to the situation in molecular aggregates \cite{spano1989superradiance}.
Finally, the restriction given in Eq. (\ref{eq:conservation}) has
two consequences in Eq. (\ref{eq:ratio}). First, $\Theta$ turns
the PL off for states to the right of the light line, and therefore
for LPs and a subset of UPs, as already discussed above. Second, for
a fixed value of in-plane momentum $\boldsymbol{k}$, there are at
most two possible values of out-of-plane photon wavevectors $k_{zd}=\pm\sqrt{\epsilon_{d}\frac{\omega_{y_{\boldsymbol{k}}}^{2}}{c^{2}}-|\boldsymbol{k}|^{2}}$,
so the rate scales as the electric field ($\propto\omega_{y_{\boldsymbol{k}}}$)
times the density of photonic states proportional to $\frac{dk_{zd}}{d\omega}|_{\omega=\omega_{y_{\boldsymbol{k}}}}=\frac{\sqrt{|\boldsymbol{k}|^{2}+k_{zd}^{2}}}{ck_{dz}}=\frac{\omega_{y_{\boldsymbol{k}}}}{ck_{dz}}$;
this frequency scaling is drastically different from the single-molecule
case, where all photon-propagation directions are available for PL.
Interestingly, there is a critical in-plane wavevector $\boldsymbol{k^{*}}=\frac{\sqrt{\epsilon_{d}}\omega_{+_{\boldsymbol{k}^{*}}}}{c}$
for which $k_{zd}=0$, which coincides with the transition between
bright and non-emissive UPs. This critical propagation direction is
along grazing incidence, where the emitted photon travels parallel
to the metal surface. At $\boldsymbol{k}^{*}$, $\frac{dk_{zd}}{d\omega}|_{\omega=\omega_{y_{\boldsymbol{k^{*}}}}}\to\infty$,
so $|\langle\gamma_{y_{\boldsymbol{k}},0}\rangle_{iso}|,\,|\langle\gamma_{y_{\boldsymbol{k}},ref}\rangle_{iso}|$
diverge, yet $r_{\boldsymbol{k^{*}}}\to-1$ (both lossless and lossy
interfaces are perfectly reflective at grazing incidence, see Fig.
\ref{fig:dispersion}). The singularities in $\langle\gamma_{y_{\boldsymbol{k}},i}\rangle_{iso}$
are van Hove anomalies \cite{van1953occurrence,andreani1991radiative,remeika_fogler}
arising from the effective one-dimensional nature of the PL process
where, given a fixed in-plane momentum $\boldsymbol{k}$ dictated
by the SP mode, the (one-dimensional) out-of-plane momentum $k_{zd}$
must be chosen to conserve energy. Incidentally, as explained in \cite{de2007colloquium},
these singularities are also at the origin of the narrow resonances
observed in surface lattice plasmons \cite{zou2004silver}, a phenomenon
described under the umbrella of Rayleigh-Wood anomalies \cite{Wood1935,Fano1941}.
The physical origin of these collective state anomalies is thus different
from single molecule emission anomalies in quasi-one-dimensional photonic
crystals \cite{viasnoff2005spontaneous}, and altogether can yield
substantial enhancements of PL rates ($\frac{\langle\gamma_{y_{\boldsymbol{k}}}\rangle_{iso}}{\gamma_{SM,0}}>10^{9}$).

\begin{figure}
\centering{}\includegraphics[scale=0.3]{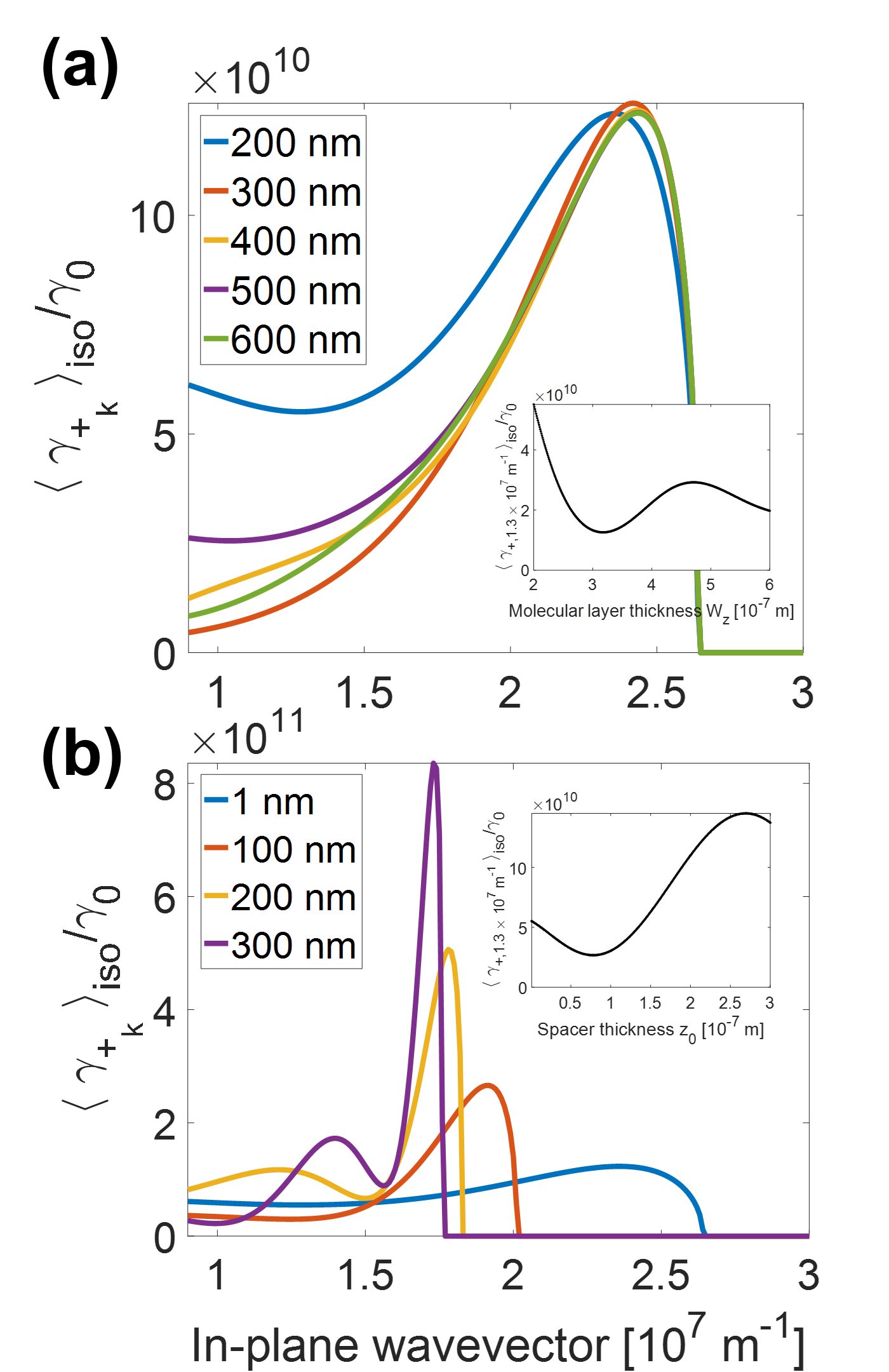}\caption{\emph{Normalized isotropic PL rates for the UP band $\frac{\langle\gamma_{+_{\boldsymbol{k}}}\rangle_{iso}}{\gamma_{SM}}$}
as a function of (a) molecular layer and (b) spacer thickness, $W_{z}$
and $z_{0}$, respectively. Results are presented for a maximal density
$\rho=3.7\times10^{10}\,\text{molecules/\ensuremath{\mu}m}^{3}$.
Insets trace the PL rate for $|+_{1.3\times10^{7}\,\text{m}^{-1}}\rangle$
which oscillates with $W_{z}$ and $z_{0}$. Oscillations in (a) are
due to interferences of the PL processes for molecules at different
heights of the layer, while oscillations in (b) are due to reflected
waves from the metal. \label{fig:thickness}}
\end{figure}

The strong-coupling theory above is a good description of the plexciton
system when disorder $\Sigma_{e}$ in the molecular excitation energies
across the layer is small. Quantitatively, we expect our theory to
fail when $\text{min}_{y}\left|\omega_{y_{\boldsymbol{k}}}-\omega_{e}\right|<\Sigma_{e}$,
where $\Sigma_{e}\approx10\,\text{meV}$ is a typical value for organic
samples \cite{jang2001characterization}. That is, for the phenomenology
described in this article, light-matter coupling must be strong enough
that the critical point (see Fig. \ref{fig:dispersion}a) must lie
above the bare exciton energy by at least $\Sigma_{e}$. For the parameters
used in Fig. \ref{fig:dispersion}, our theory does not hold for plexciton
and dark exciton states with $|\boldsymbol{k}|<0.9\times10^{7}\,\text{m}^{-1}$,
which can efficiently form superpositions of states with various $\boldsymbol{k}$
values to localize and feature $O(\gamma_{SM,0})$ PL rates. On the
other hand, it is known that vibrational relaxation dynamics between
polaritons and dark states exhibit ultrafast relaxation rates ($\sim50$
fs) \cite{agranovich2003,ribeiro2018polariton}. A detailed characterization
of vibrational relaxation processes in molecular polaritons is a complex
task on itself and beyond the scope of this work, but will be irrelevant
when PL rates become much faster, such as in our case of interest.
For the time being, these estimates suffice to place constraints on
the applicability of the present theory. An important observation
is that owing to vibrational relaxation, a much stronger PL is typically
expected from the lower polariton branch compared to the upper one
whenever the former is emissive \cite{PhysRevLett.82.3316}. Thus,
the presently described superradiant PL afforded by UPs renders plexcitonic
systems unique in terms of polariton photophysics, and could be exploited
to suppress photochemical pathways \cite{Hutchison2012} or induce
new thermodynamic equilibria \cite{Thomas2016}.

Finally, we analyze other coherent effects associated with plexciton
PL. Eq. (\ref{eq:f}) contains finite Fourier transforms of the coupling
$J_{\boldsymbol{k}}(z)$ at the wavevector $k_{zd}$, which arise
from interferences of emission pathways due to molecules at different
heights of the layer. This implies oscillatory features of $\langle\gamma_{y_{\boldsymbol{k}},i}\rangle_{iso}$
as a function of molecular layer thickness $W_{z}$ across an $O(k_{zd}^{-1})$
range of values. As explained in the previous paragraph, our theory
is valid for strong coupling, and therefore, close to the anticrossing
region and the critical wavevector $\boldsymbol{k=k^{*}}$, so $k_{dz}$
is typically small and this oscillatory effect is not appreciable
unless $W_{z}$ is very big ($>1\,\mu\text{m}$) or if $\rho$ increases.
To see this oscillatory effect, we increase the density to a maximal
value $\rho=3.7\times10^{10}\,\text{molecules/\ensuremath{\mu}}\text{m}^{3}$,
associated with the minimum van der Waals contact distance between
chromophores, and obtain Fig. \ref{fig:thickness}a. The inset shows
results for a superradiant state $|+_{1.3\times10^{7}\,\text{m}^{-1}}\rangle$,
which exhibits PL oscillations as a function of $W_{z}$ within the
first 600 nm. It is also interesting to examine yet another oscillatory
behavior of $\langle\gamma_{y_{\boldsymbol{k}},i}\rangle_{iso}$ obtained
by varying the spacer thickness $z_{0}$. This effect is due exclusively
to interference between incident and reflected waves, and therefore,
occurs when $z_{0}$ varies across photonic distances, by analogy
to the aforementioned Drexhage \cite{drexhage1968variation,drexhage1970influence}
and CPS \cite{chance1974lifetime,chance1978molecular} problem. Since
we want to stay in the strong coupling regime, we keep the maximal
$\rho$ while varying $z_{0}$ up to a threshold $\sim400\,\text{nm}$,
after which light-matter coupling becomes weak for the wavevectors
of interest. Fig. \ref{fig:thickness}b shows the oscillatory behavior
of the PL for $|+_{1.3\times10^{7}\,\text{m}^{-1}}\rangle$ as a function
of $z_{0}$.

To conclude, we have presented a comprehensive quantum formalism to
study the PL properties of plexciton systems, thus generalizing the
paradigmatic problem of single-molecule PL next to a metal interface.
We have shown there is rich phenomenology associated with superradiance
and the transition between bright and non-emissive plexciton states.
Finally, we have elucidated two types of coherent effects exhibited
by plexciton PL at high molecular densities. These include generalizations
of the well-known PL oscillations due to reflected waves from the
metal, but also new coherences that arise from multiple emission pathways
corresponding to different distances of the molecules from the metal
layer (a result of finite thickness of the molecular layer). The described
properties can be readily verified in experiments and present new
control knobs to manipulate and design photophysical properties in
molecular materials.

J.Y.Z. acknowledges discussions with Misha Fogler, Jean-Jacques Greffet,
Nicholas Rivera, Matthew Du, and Raphael Ribeiro, as well as funding
from NSF CAREER CHE 1654732 as well as UCSD startup. S.K.S. acknowledges
support from the Center for Excitonics, an Energy Frontier Research
Center funded by the U.S. Department of Energy under Award DE-SC0001088
and the Ministry of Education and Science of the Russian Federation
for supporting the research in the framework of the state assignment,
award \#3.2166.2017/4.6. V.M. acknowledges funding from U. S. Department
of Energy under Award DE-SC0017760.

\newpage{}

~

\newpage{}

 \onecolumngrid

\begin{center}
\textbf{\Large{}Supplemental Material for ``Molecular emission near
metal interfaces: the polaritonic regime''}{\Large\par}
\par\end{center}

\global\long\def\theequation{S\arabic{equation}}
 \setcounter{equation}{0}

\global\long\def\thefigure{S\arabic{figure}}
 \setcounter{figure}{0}
\begin{center}
Joel Yuen-Zhou$^{1}$, Semion K. Saikin$^{2,3}$, Vinod M. Menon$^{4}$
\par\end{center}

\begin{center}
\emph{$^{1}$Department of Chemistry and Biochemistry, University
of California San Diego, La Jolla, CA, USA.}
\par\end{center}

\begin{center}
\emph{$^{2}$Department of Chemistry and Chemical Biology, Harvard
University, Cambridge, MA, USA.}
\par\end{center}

\begin{center}
\emph{$^{3}$}\textit{Institute of Physics, Kazan Federal University,
Kazan, Russian Federation.}
\par\end{center}

\begin{center}
\emph{$^{4}$}\textit{Department of Physics, Graduate Center and City
College of New York, City University of New York, New York, New York,
USA.}
\par\end{center}

~

joelyuen@ucsd.edu

~

Section \ref{sec:SP-modes:-properties} considers the quantization
of lossless SP modes, which is utilized in the plexciton Hamiltonian
of Eq. (\ref{eq:Hk}) (derived in Section \ref{sec:Derivation-of-plexciton}).
Since PL rates are derived within a Fermi golden rule approach, neglect
of losses amounts to regarding the initial and final states in Eq.
(\ref{eq:gamma_sk}) to be eigenstates of Eq. (\ref{eq:Hk}), which
offers a good zeroth-order description of the plexciton setup in the
strong coupling regime, where the linewidth due to metal loss is much
smaller than the Rabi splitting. Derivations of PL rates within the
modal representation and the dyadic Green's function formalism are
carried out in Section \ref{sec:Derivation-of-plexciton}. Finally,
in Section \ref{sec:Effects-of-metal}, the PL expressions are shown,
within the formalism of M-QED, to agree with the case where metal
losses are taken into account in the UHP photonic density of states.

\section{SP modes: properties and quantization\label{sec:SP-modes:-properties}}

In this Section, we present the main steps to quantize lossless SP
modes. For further details, the reader might wish to consult \cite{archambault}
or our previous work \cite{yuen2016plexciton}, in particular, its
Supplementary Note 1.

\subsection{Properties of SP modes}

We summarize the main features of SP modes arising at the interface
between a dielectric material $d$ ($z>0$) and a (Drude) metal $m$
($z<0$), each with relative permittivities $\epsilon_{d}$ and $\epsilon_{m}(\omega)=\epsilon_{\infty}-\frac{\omega_{P}^{2}}{\omega^{2}+i\omega\Gamma}$,
where $\omega_{P}$ is the plasma frequency and $\Gamma$ is a damping
constant \cite{novotny,maier} (we hereafter set $\Gamma=0$, and
revisit the effects of metal losses in Section \ref{sec:Effects-of-metal}).
The modes are labeled by an \emph{in-plane} wavevector $\boldsymbol{k}$
and are characterized by the following electromagnetic fields,

\begin{subequations}\label{eq:modes}
\begin{eqnarray}
\vec{E}(\boldsymbol{k}) & = & \mathcal{A}_{\boldsymbol{k}}\boldsymbol{E}(\boldsymbol{k})e^{i(\boldsymbol{k}\cdot\boldsymbol{r}-\omega_{\boldsymbol{k}}^{SP}t)},\label{eq:mode_E}\\
\vec{B}(\boldsymbol{k}) & = & \mathcal{A}_{\boldsymbol{k}}\boldsymbol{B}(\boldsymbol{k})e^{i(\boldsymbol{k}\cdot\boldsymbol{r}-\omega_{\boldsymbol{k}}^{SP}t)}.\label{eq:mode_B}
\end{eqnarray}

\end{subequations}\noindent where $\mathcal{A}_{\boldsymbol{k}}$
is a mode amplitude (see Section \ref{subsec:Quantization}). The
frequencies corresponding to each of these modes are $\omega_{\boldsymbol{k}}^{SP}=c\Omega_{\boldsymbol{k}}^{SP}$,
where,

\begin{eqnarray}
\Omega_{\boldsymbol{k}}^{SP} & = & k\sqrt{\frac{\epsilon_{d}+\epsilon_{m}}{\epsilon_{d}\epsilon_{m}}}\nonumber \\
 & = & \sqrt{\frac{\epsilon_{d}\Omega_{P}^{2}+k^{2}(\epsilon_{d}+\epsilon_{\infty})-\sqrt{[\epsilon_{d}\Omega_{P}^{2}+k^{2}(\epsilon_{d}+\epsilon_{\infty})]^{2}-4\epsilon_{\infty}\epsilon_{d}k^{2}\Omega_{P}^{2}}}{2\epsilon_{\infty}\epsilon_{d}}},\label{eq:SP_disp}
\end{eqnarray}
This equation was used to plot the SP dispersion band in Fig. \ref{fig:dispersion}a.
Another way to write it is,

\begin{equation}
\Omega_{\boldsymbol{k}}^{SP}=\frac{\sqrt{|\boldsymbol{k}|^{2}+k_{zd}^{2}}}{\sqrt{\epsilon_{d}}}=\frac{\sqrt{|\boldsymbol{k}|^{2}+k_{zm}^{2}}}{\sqrt{\epsilon_{m}}}.\label{eq:dispersion_Omega}
\end{equation}
Eq. (\ref{eq:dispersion_Omega}) implies that for SP modes, the vertical
components of the wavevector in the dielectric ($k_{zd}$) and the
metal ($k_{zm}$) depend on $|\boldsymbol{k}|=k$. Since they represent
evanescent decays, we write $k_{zd}=i\alpha_{dk}$ and $k_{zm}=-i\alpha_{mk}$,
where $\alpha_{dk},\alpha_{mk}\in\mathcal{\Re}$ and positive. The
vector components of the modes in Eq. \ref{eq:modes} are given by,

\begin{subequations}\label{eq:modes1}

\begin{eqnarray}
\boldsymbol{E} & = & \mathcal{A}_{\boldsymbol{k}}[\Theta(-z)\boldsymbol{E}_{m}e^{\alpha_{mk}z}+\Theta(z)\boldsymbol{E}_{d}e^{-\alpha_{dk}z}],\label{eq:totalE-1}\\
\boldsymbol{B} & = & \mathcal{A}_{\boldsymbol{k}}[\Theta(-z)\boldsymbol{B}_{m}e^{\alpha_{mk}z}+\Theta(z)\boldsymbol{B}_{d}e^{-\alpha_{dk}z}],\label{eq:totalB-1}
\end{eqnarray}
\end{subequations}\noindent and each of the involved vectors are
given by

\begin{subequations}\label{eq:SPmodes}

\begin{eqnarray}
\boldsymbol{E}_{d} & = & (1,0,\frac{ik}{\alpha_{dk}}),\label{eq:Ed}\\
\boldsymbol{B}_{d} & = & (0,-\frac{i\omega_{\boldsymbol{k}}^{SP}\epsilon_{d}}{\alpha_{dk}c^{2}},0)\label{eq:Bd}
\end{eqnarray}
and

\begin{eqnarray}
\boldsymbol{E}_{m} & = & (1,0,-\frac{ik}{\alpha_{mk}}),\label{eq:Em}\\
\boldsymbol{B}_{m} & = & (0,\frac{i\omega_{\boldsymbol{k}}^{SP}\epsilon_{m}}{\alpha_{mk}c^{2}},0).\label{eq:Bm}
\end{eqnarray}

\end{subequations}\noindent where the Cartesian notation we are using
in this Section is expressed using the unit vectors $\hat{\boldsymbol{k}},\hat{\boldsymbol{\theta}}_{\boldsymbol{k}},\hat{\boldsymbol{z}}$
such that $\hat{\boldsymbol{k}}\times\hat{\boldsymbol{\theta}}_{\boldsymbol{k}}=\hat{\boldsymbol{z}}$,
so that $\boldsymbol{E}=(E_{k},E_{\theta_{k}},E_{z})$, where $E_{i}=\boldsymbol{E}\cdot\hat{\boldsymbol{i}}$
(note that the tangential direction of $\hat{\boldsymbol{\theta}}_{\boldsymbol{k}}$
is with respect to $\hat{\boldsymbol{k}}$ and not to $\hat{\boldsymbol{r}}$).
Beware that we are using a different Cartesian notation than the one
in the main text.

\subsection{Quantization of SP modes\label{subsec:Quantization}}

The energy in the $\boldsymbol{k}$th mode in Eq. (\ref{eq:modes})
is quadratic in the fields \cite{novotny,archambault},

\begin{equation}
H_{SP,\boldsymbol{k}}=\frac{1}{2}\sum_{i}\int dV\Bigg[\epsilon_{0}\sum_{j}\frac{d(\omega\epsilon^{*}(\omega))}{d\omega}|(\vec{E})_{j}|^{2}+\frac{1}{\mu_{0}\mu}|(\vec{B})_{i}|^{2}\Bigg]|\mathcal{A}_{\boldsymbol{k}}|^{2},\label{eq:energy_sp}
\end{equation}
where $i,j\in\{k,\theta_{k},z\}$. Taking the integration volume to
be a box of in-plane area $S$ and infinite height, we plug in Eq.
\ref{eq:SPmodes} into Eq. (\ref{eq:energy_sp}) and use $\int dxdy=S$,
$\int_{-\infty}^{0}dze^{-2\alpha_{d}z}=\frac{1}{2\alpha_{d}}$, and
$\int_{-\infty}^{0}dze^{-2\alpha_{m}z}=\frac{1}{2\alpha_{m}}$ to
obtain,

\begin{eqnarray}
H_{SP,\boldsymbol{k}} & = & S|\mathcal{A}_{\boldsymbol{k}}|^{2}\sum_{i}\Bigg\{\Bigg[\epsilon_{0}\epsilon_{d}|E_{i,d}|^{2}+\frac{1}{\mu_{0}\mu}|B_{i,d}|^{2}\Bigg]\frac{1}{4\alpha_{dk}}+\Bigg[\epsilon_{0}\frac{d(\omega\epsilon_{m}(\omega))}{d\omega}|E_{i,m}|^{2}+\frac{1}{\mu_{0}\mu}|B_{i,m}|^{2}\Bigg]\frac{1}{4\alpha_{mk}}\Bigg\}\nonumber \\
 & = & S\frac{\epsilon_{0}L_{\boldsymbol{k}}}{4}(\mathcal{A}_{\boldsymbol{k}}\mathcal{A}_{\boldsymbol{k}}^{*}+\mathcal{A}_{\boldsymbol{k}}^{*}\mathcal{A}_{\boldsymbol{k}}),\label{eq:H_SP_k_supp}
\end{eqnarray}
where the mode length $L_{\boldsymbol{k}}$ \cite{archambault,yuen2016plexciton}
is defined as,

\begin{eqnarray}
L_{\boldsymbol{k}} & = & \sum_{i}\Bigg\{\Bigg[\epsilon_{0}\epsilon_{d}|E_{i,d}|^{2}+\frac{1}{\mu_{0}\mu}|B_{i,d}|^{2}\Bigg]\frac{1}{2\epsilon_{0}\alpha_{dk}}+\Bigg[\epsilon_{0}\frac{d(\omega\epsilon_{m}(\omega))}{d\omega}\Bigg|_{\omega=\frac{\Omega_{0}}{c}}|E_{i,m}|^{2}+\frac{1}{\mu_{0}\mu}|B_{i,m}|^{2}\Bigg]\frac{1}{2\epsilon_{0}\alpha_{mk}}\Bigg\}\nonumber \\
 & = & \frac{-\epsilon_{m}}{\alpha_{dk}}+\frac{1}{2\alpha_{mk}}\Bigg[\frac{d(\omega\epsilon_{m}(\omega))}{d\omega}\Bigg|_{\omega=\frac{\Omega(\boldsymbol{k})}{c}}\Bigg(\frac{\epsilon_{m}-\epsilon_{d}}{\epsilon_{m}}\Bigg)-\epsilon_{m}-\epsilon_{d}\Bigg].\label{eq:Lk_calculation}
\end{eqnarray}
In this derivation, we have used $\mu=1$, $\epsilon_{0}\mu_{0}=c^{-2}$
as well as Eqs. (\ref{eq:SP_disp}) and (\ref{eq:dispersion_Omega})\footnote{The final result in Eq. (\ref{eq:Lk_calculation}) is twice what is
reported in the Supplementary Material of \cite{archambault}, although
we believe our result here \cite{yuen2016plexciton} is correct.}.

Eq. (\ref{eq:H_SP_k_supp}) is quadratic in the electric field amplitude
$\mathcal{A}_{\boldsymbol{k}}$, so each $\boldsymbol{k}$ mode corresponds
to a harmonic oscillator,

\begin{equation}
H_{SP,\boldsymbol{k}}=\frac{\hbar\omega_{\boldsymbol{k}}^{SP}}{2}(\zeta_{\boldsymbol{k}}\zeta_{\boldsymbol{k}}^{*}+\zeta_{\boldsymbol{k}}^{*}\zeta_{\boldsymbol{k}}),\label{eq:quantization}
\end{equation}
so that the amplitude $\mathcal{A}_{\boldsymbol{k}}$ can be related
to $\zeta_{\boldsymbol{k}}$ by,

\begin{eqnarray}
\mathcal{A}_{\boldsymbol{k}} & = & \sqrt{\frac{2\hbar\omega_{\boldsymbol{k}}^{SP}}{\epsilon_{0}SL_{\boldsymbol{k}}}}\zeta_{\boldsymbol{k}}.\label{eq:A_in_terms_of_alpha}
\end{eqnarray}
Promoting the complex amplitudes to SP annhilation and creation operators,
$\zeta_{\boldsymbol{k}}\to a_{\boldsymbol{k}}$ and $\zeta_{\boldsymbol{k}}^{*}\to a_{\boldsymbol{k}}^{\dagger}$
with $[a_{\boldsymbol{k}},a_{\boldsymbol{k}}^{\dagger}]=1$,

\begin{eqnarray}
H_{SP,\boldsymbol{k}} & = & \frac{\hbar\omega_{\boldsymbol{k}}^{SP}}{2}(a_{\boldsymbol{k}}a_{\boldsymbol{k}}^{\dagger}+a_{\boldsymbol{k}}^{\dagger}a_{\boldsymbol{k}})\nonumber \\
 & = & \frac{\hbar\omega_{\boldsymbol{k}}^{SP}}{2}\Big(a_{\boldsymbol{k}}^{\dagger}a_{\boldsymbol{k}}+\frac{1}{2}\Big).\label{eq:quantization_final}
\end{eqnarray}
The Hamiltonian $H_{SP}=\sum_{\boldsymbol{k}}H_{SP,\boldsymbol{k}}-\sum_{\boldsymbol{k}}\frac{\hbar\omega_{\boldsymbol{k}}^{SP}}{2}=\hbar\omega_{\boldsymbol{k}}^{SP}a_{\boldsymbol{k}}^{\dagger}a_{\boldsymbol{k}}$
in the main text ignores the background zero-point energies of the
SP modes since the energy $\omega_{e}$ of the excitons already takes
this background into account.

The electric field and magnetic induction are superpositions of amplitudes
in each of the $\boldsymbol{k}$ modes,

\begin{subequations}\label{eq:e_m_fields}

\begin{eqnarray}
\vec{\mathcal{E}} & = & \sum_{\boldsymbol{k}}\mathcal{A}_{\boldsymbol{k}}\vec{E}(\boldsymbol{k}).\label{eq:superposition_E}\\
\vec{\mathcal{B}} & = & \sum_{\boldsymbol{k}}\mathcal{A}_{\boldsymbol{k}}\vec{B}(\boldsymbol{k}).\label{eq:superposition_B}
\end{eqnarray}
\end{subequations}Promoting these amplitudes to quantum operators
in the Heisenberg picture, and using Eq. (\ref{eq:A_in_terms_of_alpha}),

\begin{subequations}

\begin{eqnarray}
\hat{\boldsymbol{\mathcal{E}}}(\boldsymbol{r},t) & = & \sum_{\boldsymbol{k}}\sqrt{\frac{\hbar\omega_{\boldsymbol{k}}^{SP}}{2\epsilon_{0}SL_{\boldsymbol{k}}}}a_{\boldsymbol{k}}\vec{E}(\boldsymbol{k})+\sqrt{\frac{\hbar\omega_{\boldsymbol{k}}^{SP}}{2\epsilon_{0}SL_{\boldsymbol{k}}}}a_{\boldsymbol{k}}^{\dagger}\vec{E}^{*}(\boldsymbol{k}),\label{eq:operator_E}\\
\hat{\boldsymbol{\mathcal{B}}}(\boldsymbol{r},t) & = & \sum_{\boldsymbol{k}}\sqrt{\frac{\hbar\omega_{\boldsymbol{k}}^{SP}}{2\epsilon_{0}SL_{\boldsymbol{k}}}}a_{\boldsymbol{k}}\vec{B}(\boldsymbol{k})+\sqrt{\frac{\hbar\omega_{\boldsymbol{k}}^{SP}}{2\epsilon_{0}SL_{\boldsymbol{k}}}}a_{\boldsymbol{k}}^{\dagger}\vec{B}^{*}(\boldsymbol{k}),\label{eq:operator_B}
\end{eqnarray}
\end{subequations}where we have used the fact that $\boldsymbol{\mathcal{E}}$
and $\mathcal{\boldsymbol{B}}$ are real valued to write the results
in a symmetrized fashion.

In the main text, the electric field in the Schrodinger picture for
the dielectric region $z>0$ is,

\begin{eqnarray}
\hat{\boldsymbol{\mathcal{E}}}(\boldsymbol{r}_{\boldsymbol{n}s}) & = & \sum_{\boldsymbol{k}}\sqrt{\frac{\hbar\omega_{\boldsymbol{k}}^{SP}}{2\epsilon_{0}SL_{\boldsymbol{k}}}}a_{\boldsymbol{k}}\underbrace{(1,0,\frac{ik}{\alpha_{dk}})}_{=\boldsymbol{E}_{\boldsymbol{k}}=\hat{\boldsymbol{k}}+\frac{i|\boldsymbol{k}|}{\alpha_{dk}}\hat{\boldsymbol{z}}}e^{i\boldsymbol{k}\cdot\boldsymbol{r}_{\boldsymbol{n}}-\alpha_{d}z_{s}}+\mbox{h.c.}\label{eq:operator_E_schrodinger}
\end{eqnarray}

\section{Derivation of plexciton Hamiltonian, Eq. (\ref{eq:Hk})\label{sec:Derivation-of-plexciton}}

In this Section, we abound on the quantum-mechanical formalism to
describe the plexciton system and ultimately leading to Eq. (\ref{eq:Hk})
in the main text. The Hamiltonian $H$ for this setup of interest
can be expressed as

\begin{equation}
H=H_{exc}+H_{SP}+H_{exc-SP},\label{eq:total_hamiltonian}
\end{equation}
where $H_{exc}=\sum_{\boldsymbol{n},s}\hbar\omega_{\boldsymbol{n}s}^{(e)}\sigma_{\boldsymbol{n}s}^{\dagger}\sigma_{\boldsymbol{n}s}$,
$H_{SP}=\sum_{\boldsymbol{k}}H_{\boldsymbol{k}}=\sum_{\boldsymbol{k}}\hbar\omega_{\boldsymbol{k}}^{SP}a_{\boldsymbol{k}}^{\dagger}a_{\boldsymbol{k}},$
and $H_{exc-SP}=-\sum_{\boldsymbol{n},s}\hat{\boldsymbol{\mu}}_{\boldsymbol{n}s}\cdot\hat{\boldsymbol{\mathcal{E}}}(\boldsymbol{r}_{\boldsymbol{n}s})$
denote the energetics of the molecular exciton layer, the SP modes
of the metal, and the dipolar coupling between them; we neglect dipole-dipole
hopping interactions between exciton states, since they play a minor
role in our problem compared to light-matter coupling. We consider
a layer containing $N=N_{x}N_{y}N_{z}$ emitters, with $N_{i}$ molecules
along each axis $i$; $\boldsymbol{n}=(n_{x},n_{y})$ is a Cartesian
pair of integers denoting the in-plane location $\boldsymbol{r}_{\boldsymbol{n}}=\boldsymbol{n}\cdot(\Delta_{x},\Delta_{y})$
of a chromophore, while the integer $s$ denotes its vertical location
$z_{s}\hat{\boldsymbol{z}}$ at $z_{s}=z_{0}+s\Delta_{z}$; $z=z_{0}>0$
and $z=z_{f}$ label the lowest and highest vertical coordinate for
a molecule in the layer and $z=0$ refers to the metal-dielectric
spacer interface. $W_{z}=N_{z}\Delta_{z}$ is the vertical thickness
of the molecular layer. With this convention, the location of the
chromophore labeled $\boldsymbol{n}s$ is $\boldsymbol{r}_{\boldsymbol{n}s}=\boldsymbol{r}_{n}+(z_{0}+s\Delta_{z})\hat{\boldsymbol{z}}$;
its corresponding single(-Frenkel)-exciton energy and creation (annhilation)
operator are $\hbar\omega_{\boldsymbol{n}s}$ and $\sigma_{\boldsymbol{n}s}^{\dagger}$($\sigma_{\boldsymbol{n}s})$.
Thus, the dipole operator reads as $\hat{\boldsymbol{\mu}}_{\boldsymbol{n}s}=\boldsymbol{\mu}_{\boldsymbol{n}s}(\sigma_{\boldsymbol{n}s}^{\dagger}+\sigma_{\boldsymbol{n}s})$.
The parameters for $H_{SP}$ and $\hat{\boldsymbol{\mathcal{E}}}(\boldsymbol{r}_{\boldsymbol{n}s})$
have been defined in the previous Section.

For simplicity, we first consider identical exciton transitions ($\omega_{\boldsymbol{n}s}=\omega_{e}$,
$\boldsymbol{\mu}_{\boldsymbol{n}s}=\boldsymbol{\mu}_{eg}$). This
is a good approximation because we are mainly interested in the ``anticrossing''
region, where light and matter couple much more strongly than the
energetic disorder due to imperfections of the molecular layer; the
extension to an isotropically oriented sample is straightforward \cite{gonzalez_tudela}.
We also simplify the problem by invoking the rotating-wave approximation
(ignore ``off-resonant'' $\sigma_{\boldsymbol{n}s}^{\dagger}a_{\boldsymbol{k}}^{\dagger}$
and $\sigma_{\boldsymbol{n}s}a_{\boldsymbol{k}}$ terms in $H_{exc-SP}$)
\cite{Scully1997}. Given these assumptions, translational invariance
allows for a Bloch decomposition of the Hamiltonian in Eq. (\ref{eq:total_hamiltonian})
as $H=\sum_{\boldsymbol{k}}H_{\boldsymbol{k}}$, where $H_{\boldsymbol{k}}$
is given by Eq. (\ref{eq:Hk}). In that expression, we have defined
$N_{x}N_{y}$ collective $\boldsymbol{k}=(k_{x},k_{y})$ exciton states
($k_{i}=\frac{2\pi m_{i}}{N_{i}\Delta_{i}}$ for $m_{i}=-\frac{N_{i}}{2},\cdots,\frac{N_{i}}{2}-1$)
represented by the operators $\sigma_{\boldsymbol{k}}^{\dagger}=\sqrt{\frac{N_{x}N_{y}}{S}}\frac{\sum_{s}J_{\boldsymbol{k}}(z_{s})}{\mathcal{J}_{\boldsymbol{k}}}\sigma_{\boldsymbol{k}s}^{\dagger}$
\cite{gonzalez_tudela}, where $\sigma_{\boldsymbol{k}s}^{\dagger}=\frac{\sum_{\boldsymbol{n}}\sigma_{\boldsymbol{n}s}^{\dagger}e^{i\boldsymbol{k}\cdot\boldsymbol{r}_{\boldsymbol{n}}}}{\sqrt{N_{x}N_{y}}}$,
$J_{\boldsymbol{k}}(z)=\sqrt{S}\left(\sqrt{\frac{\hbar\omega_{\boldsymbol{k}}^{SP}}{2\epsilon_{0}SL_{\boldsymbol{k}}}}\boldsymbol{\mu}_{eg}\cdot\boldsymbol{E}_{\boldsymbol{k}}e^{-\alpha_{\boldsymbol{k}}z}\right)$,
and $\mathcal{J}_{\boldsymbol{k}}=\sqrt{\rho\int_{z_{0}}^{z_{f}}dz|J_{\boldsymbol{k}}(z)|^{2}}$
is an effective collective coupling, with $\rho$ being the density
of molecules in the layer. It follows that we can write $\sigma_{\boldsymbol{k}}^{\dagger}=\sum_{\boldsymbol{n},s}d_{\boldsymbol{n}s}\sigma_{\boldsymbol{n}s}^{\dagger}$,
where
\begin{equation}
d_{\boldsymbol{n}s}=\sqrt{\frac{1}{S}}\frac{\sum_{s}J_{\boldsymbol{k}}(z_{s})}{\mathcal{J}_{\boldsymbol{k}}}e^{i\boldsymbol{k}\cdot\boldsymbol{r}_{\boldsymbol{n}}}.\label{eq:dns}
\end{equation}
$H_{dark,\boldsymbol{k}}=\hbar\omega_{e}(\mathbb{I}_{exc,\boldsymbol{k}}-\sigma_{\boldsymbol{k}}^{\dagger}\sigma_{\boldsymbol{k}}$)
describes the energetics of the manifold of dark exciton states which
do not couple to the SP fields and which contains $N_{x}N_{y}(N_{z}-1)$
exciton states (so that the original number of exciton states is conserved).
$\mathbb{I}_{exc,\boldsymbol{k}}$ is the identity operator on the
Hilbert space defined by $\{\sigma_{\boldsymbol{k}s}^{\dagger}\}_{s=1}^{N_{z}}$.
Finally, $H_{umklapp,\boldsymbol{k}}=\sum_{\boldsymbol{q}=(\frac{2\pi q_{x}}{\Delta_{x}},\frac{2\pi q_{y}}{\Delta_{y}})\neq0}\Big[\hbar\omega_{\boldsymbol{k}+\boldsymbol{q}}^{SP}a_{\boldsymbol{k}+\boldsymbol{q}}^{\dagger}a_{\boldsymbol{k}+\boldsymbol{q}}+(\mathcal{J}_{\boldsymbol{k}+\boldsymbol{q}}\sigma_{\boldsymbol{k}}^{\dagger}a_{\boldsymbol{k}+\boldsymbol{q}}+\mbox{h.c.})\Big]$,
where $\boldsymbol{q}$ is a vector of integers, reflects the fact
that excitons with a wavevector $\boldsymbol{k}$ can couple to an
infinite countable set of SP modes beyond the first Brillouin zone.
For our purposes, $H_{umklapp,\boldsymbol{k}}$ can be ignored, as
its associated SP modes are far off-resonant with respect to the exciton
states.

A straightforward way to generalize this result to an isotropic sample
is to assume that each location $r_{\boldsymbol{n}s}$ contains three
equivalent excitonic transitions with transition dipole amplitudes
$\frac{|\boldsymbol{\mu}_{eg}|}{\sqrt{3}}$ labeled by $\sigma_{\boldsymbol{n}s,\hat{\boldsymbol{j}}}^{\dagger}$
along the $j=x,y,z$ directions. The results above can be recycled
provided that we define $\sigma_{\boldsymbol{n}s}^{\dagger}=\frac{\sum_{j}(E_{\boldsymbol{k}}\cdot\hat{\boldsymbol{j}})\sigma_{\boldsymbol{n}s,j}^{\dagger}}{\sqrt{\sum_{j}|E_{\boldsymbol{k}}\cdot\hat{\boldsymbol{j}}|^{2}}}$
and adapt our definition $J_{\boldsymbol{k}}(z)=\sqrt{S}\left(\sqrt{\frac{\hbar\omega_{\boldsymbol{k}}^{SP}}{2\epsilon_{0}SL_{\boldsymbol{k}}}}|\boldsymbol{\mu}_{eg}|\frac{\sum_{j}\hat{\boldsymbol{j}}}{\sqrt{3}}\cdot\boldsymbol{E}_{\boldsymbol{k}}e^{-\alpha_{\boldsymbol{k}}z}\right)$
upon which $\mathcal{J}_{\boldsymbol{k}}=\frac{\mathcal{J}_{\boldsymbol{k}}(\boldsymbol{E}_{\boldsymbol{k}}=\hat{\boldsymbol{k}})+\mathcal{J}_{\boldsymbol{k}}(\boldsymbol{E}_{\boldsymbol{k}}=\frac{i|\boldsymbol{k}|}{\alpha_{dk}}\hat{\boldsymbol{z}})}{3}$\footnote{(Notice that the expression for the isotropic collective coupling
reported in \cite{gonzalez_tudela} (two paragraphs after Eq. 4, it
reads $g_{iso}^{2}=\frac{2}{3}g_{||}^{2}+\frac{1}{3}g_{\perp}^{2}$)
is incorrect; it should read $g_{iso}^{2}=\frac{1}{3}g_{||}^{2}+\frac{1}{3}g_{\perp}^{2}$).}.

\section{Evaluation of plexciton PL rate\label{sec:Evaluation-of-plexciton}}

\subsection{Fermi's golden rule calculation}

By defining $\boldsymbol{K}=\boldsymbol{K}_{\perp}+K_{zd}\hat{\boldsymbol{z}}$,
where $K_{zd}=\boldsymbol{K}\cdot\hat{\boldsymbol{z}}\hat{\boldsymbol{z}}$
and indexing the sums over UHP modes as $\sum_{\boldsymbol{K},\chi}=\sum_{\boldsymbol{K}_{\perp},\chi}\sum_{K_{zd}}$,
we rewrite Eq. (\ref{eq:E_UHP}) as $\hat{\boldsymbol{\mathcal{E}}}_{UHP}(\boldsymbol{r}_{\boldsymbol{n}s})=\hat{\boldsymbol{\mathcal{E}}}_{0}(\boldsymbol{r}_{\boldsymbol{n}s})+\hat{\boldsymbol{\mathcal{E}}}_{ref}(\boldsymbol{r}_{\boldsymbol{n}s})$,
where

\begin{subequations}\label{eq:E_UHP_partition}

\begin{align}
\hat{\boldsymbol{\mathcal{E}}}_{0}(\boldsymbol{r}) & =\sum_{\boldsymbol{K}_{\perp},\chi}\,\,\,\sum_{K_{zd}}\,\,\,\Theta(-\boldsymbol{K}\cdot\hat{\boldsymbol{z}})\Bigg[\frac{(b_{\boldsymbol{K},\chi}+b_{\boldsymbol{K}_{\perp}+K_{zd}\hat{\boldsymbol{z}},\chi})}{\sqrt{1+|r_{\boldsymbol{K}\chi}|^{2}}}\sqrt{\frac{\hbar\omega_{\boldsymbol{K}}^{UHP}}{2\epsilon_{0}\epsilon_{d}V}}\boldsymbol{e}_{\boldsymbol{K},\chi}e^{i\boldsymbol{K}\cdot\boldsymbol{r}}\Bigg]+\mbox{h.c.},\label{eq:E_0}\\
\hat{\boldsymbol{\mathcal{E}}}_{ref}(\boldsymbol{r}) & =\sum_{\boldsymbol{K}_{\perp},\chi}\,\,\,\sum_{K_{zd}}\,\,\,\Theta(-\boldsymbol{K}\cdot\hat{\boldsymbol{z}})\left[\frac{(b_{\boldsymbol{K},\chi}+b_{\boldsymbol{K}_{\perp}+K_{zd}\hat{\boldsymbol{z}},\chi})r_{\boldsymbol{K},\chi}}{\sqrt{1+|r_{\boldsymbol{K}\chi}|^{2}}}\sqrt{\frac{\hbar\omega_{\boldsymbol{K}}^{UHP}}{2\epsilon_{0}\epsilon_{d}V}}e^{i\boldsymbol{K}_{\perp}\cdot\boldsymbol{r}-iK_{zd}\boldsymbol{r}\cdot\hat{\boldsymbol{z}}}\boldsymbol{e}_{\boldsymbol{K}_{\perp}+K_{zd}\hat{\boldsymbol{z}},\chi}\right]+\text{h.c.}\label{eq:E_ref}
\end{align}

\end{subequations}\noindent The Fermi golden rule expression in Eq.
(\ref{eq:gamma_sk}) can be naturally partitioned as,
\begin{equation}
\gamma_{s_{\boldsymbol{k}}}=\gamma_{s_{\boldsymbol{k}},0}+\gamma_{s_{\boldsymbol{k}},ref},\label{eq:partition_of_gamma}
\end{equation}
respectively denoting incoherent contributions independently due to
the free-space and reflected waves, and a coherent term between them.
Specifically,

\begin{subequations}\label{eq:gamma_partition}

\begin{eqnarray}
\gamma_{y_{\boldsymbol{k}},0} & = & 2\times\frac{2\pi}{\hbar}\sum_{\boldsymbol{K}_{\perp},\chi}\sum_{K_{zd}<0}\Bigg[|\langle\mbox{vac};(\boldsymbol{K},\chi)_{UHP}|-\sum_{\boldsymbol{n}s}\hat{\boldsymbol{\mu}}_{\boldsymbol{n}s}\cdot\hat{\boldsymbol{\mathcal{E}}}_{0}(\boldsymbol{r}_{\boldsymbol{n}s})|y_{\boldsymbol{k}};0_{UHP}\rangle|^{2}\nonumber \\
 &  & \,\,\,\,\,\,\,\,\,\,\,\,\,\,\,\,\,\,\,\,\,\,+|\langle\mbox{vac};(\boldsymbol{K},\chi)_{UHP}|-\sum_{\boldsymbol{n}s}\hat{\boldsymbol{\mu}}_{\boldsymbol{n}s}\cdot\hat{\boldsymbol{\mathcal{E}}}_{ref}(\boldsymbol{r}_{\boldsymbol{n}s})|y_{\boldsymbol{k}};0_{UHP}\rangle|^{2}\Bigg]\delta(\hbar\omega_{y_{\boldsymbol{k}}}-\hbar\omega_{\boldsymbol{K}}^{UHP}),\label{eq:gamma_0}\\
\gamma_{y_{\boldsymbol{k}},ref} & = & 2\times\frac{2\pi}{\hbar}\sum_{\boldsymbol{K}_{\perp},\chi}\sum_{K_{zd}<0}2\Re\Bigg[\langle y_{\boldsymbol{k}};0_{UHP}|\sum_{\boldsymbol{n}s}\hat{\boldsymbol{\mu}}_{\boldsymbol{n}s}\cdot\hat{\boldsymbol{\mathcal{E}}}_{ref}(\boldsymbol{r}_{\boldsymbol{n}s})|\mbox{vac};(\boldsymbol{K},\chi)_{UHP}\rangle\nonumber \\
 &  & \times\langle\mbox{vac};(\boldsymbol{K},\chi)_{UHP}|\sum_{\boldsymbol{n'}s'}\hat{\boldsymbol{\mu}}_{\boldsymbol{n}'s'}\cdot\hat{\boldsymbol{\mathcal{E}}}_{0}(\boldsymbol{r}_{\boldsymbol{n}'s'})|y_{\boldsymbol{k}};0_{UHP}\rangle\Bigg]\delta(\hbar\omega_{y_{\boldsymbol{k}}}-\hbar\omega_{\boldsymbol{K}}^{UHP}),\label{eq:gamma_ref}
\end{eqnarray}

\end{subequations}\noindent where we have recognized the symmetry
$\sum_{K_{zd}}=2\sum_{K_{zd}<0}$. To evaluate the expressions above,
the following matrix element for $K_{zd}<0$ is handy,

\begin{eqnarray}
 &  & \langle\mbox{vac};(\boldsymbol{K},\chi)_{UHP}|-\sum_{\boldsymbol{n}s}\hat{\boldsymbol{\mu}}_{\boldsymbol{n}s}\cdot\hat{\boldsymbol{\mathcal{E}}}_{0}(\boldsymbol{r}_{\boldsymbol{n}s})|y_{\boldsymbol{k}};0_{UHP}\rangle\nonumber \\
 & = & -\zeta_{y_{\boldsymbol{k}}}^{(exc)}\langle\mbox{vac};(\boldsymbol{K},\chi)_{UHP}|\sum_{\boldsymbol{n}s}\hat{\boldsymbol{\mu}}_{\boldsymbol{n}s}\cdot\hat{\boldsymbol{\mathcal{E}}}_{\begin{array}{c}
0\end{array}}(\boldsymbol{r}_{\boldsymbol{n}s})d_{\boldsymbol{n}s}\sigma_{\boldsymbol{n}s}^{\dagger}|\mbox{vac};0_{UHP}\rangle\nonumber \\
 & = & -\frac{\zeta_{y_{\boldsymbol{k}}}^{(exc)}}{\sqrt{1+|r_{\boldsymbol{K}\chi}|^{2}}}\sqrt{\frac{\hbar\omega_{\boldsymbol{K}}^{UHP}}{2\epsilon_{0}\epsilon_{d}V}}\boldsymbol{\mu}_{ge}\cdot\boldsymbol{e}_{\boldsymbol{K},\chi}^{*}\sum_{\boldsymbol{n}s}e^{-i\boldsymbol{K}_{\perp}\cdot\boldsymbol{r}_{\boldsymbol{n}}}e^{-iK_{zd}z_{s}}\sqrt{\frac{1}{S}}\frac{J_{\boldsymbol{k}}(z_{s})}{\mathcal{J}_{\boldsymbol{k}}}e^{i\boldsymbol{k}\cdot\boldsymbol{r}_{\boldsymbol{n}}}\nonumber \\
 & = & -\frac{\zeta_{y_{\boldsymbol{k}}}^{(exc)}}{\sqrt{1+|r_{\boldsymbol{K}\chi}|^{2}}}\sqrt{\frac{\hbar\omega_{\boldsymbol{K}}^{UHP}}{2\epsilon_{0}\epsilon_{d}V}}\boldsymbol{\mu}_{ge}\cdot\boldsymbol{e}_{\boldsymbol{K},\chi}^{*}\sum_{\boldsymbol{n}s}e^{-i\boldsymbol{K}_{\perp}\cdot\boldsymbol{r}_{\boldsymbol{n}}}e^{-iK_{zd}z_{s}}\sqrt{\frac{1}{S}}\frac{J_{\boldsymbol{k}}(z_{s})}{\mathcal{J}_{\boldsymbol{k}}}e^{i\boldsymbol{k}\cdot\boldsymbol{r}_{\boldsymbol{n}}}\nonumber \\
 & = & -\frac{\zeta_{y_{\boldsymbol{k}}}^{(exc)}}{\sqrt{1+|r_{\boldsymbol{K}\chi}|^{2}}}\sqrt{\frac{\hbar\omega_{\boldsymbol{K}}^{UHP}}{2\epsilon_{0}\epsilon_{d}V}}\boldsymbol{\mu}_{ge}\cdot\boldsymbol{e}_{\boldsymbol{K},\chi}^{*}N_{x}N_{y}\delta_{\boldsymbol{K}_{\perp},\boldsymbol{k}}\sqrt{\frac{1}{S}}\frac{\frac{N_{z}}{W_{z}}\int_{z_{0}}^{z_{f}}dzJ_{\boldsymbol{k}}(z)e^{-iK_{zd}z}}{\sqrt{\rho\int_{z_{0}}^{z_{f}}dz|J_{\boldsymbol{k}}(z)|^{2}}},\label{eq:matrix_element-1}
\end{eqnarray}
where we have approximated the sum over layers by an integral $\sum_{s}\to\frac{N_{z}}{W_{z}}\int_{z_{0}}^{z_{f}}dz$
. Proceeding similarly with the reflected waves, we have for $K_{dz}<0$,

\begin{align}
 & \langle\mbox{vac};(\boldsymbol{K},\chi)_{UHP}|-\sum_{\boldsymbol{n}s}\hat{\boldsymbol{\mu}}_{\boldsymbol{n}s}\cdot\hat{\boldsymbol{\mathcal{E}}}_{ref}(\boldsymbol{r}_{\boldsymbol{n}s})|y_{\boldsymbol{k}};0_{UHP}\rangle\nonumber \\
= & -\frac{r_{\boldsymbol{K}\chi}^{*}\zeta_{y_{\boldsymbol{k}}}^{(exc)}}{\sqrt{1+|r_{\boldsymbol{K}\chi}|^{2}}}\sqrt{\frac{\hbar\omega_{\boldsymbol{K}}^{UHP}}{2\epsilon_{0}\epsilon_{d}V}}\boldsymbol{\mu}_{ge}\cdot\boldsymbol{e}_{\boldsymbol{K}_{\perp}-K_{zd}\hat{\boldsymbol{z}},\chi}^{*}N_{x}N_{y}\delta_{\boldsymbol{K}_{\perp},\boldsymbol{k}}\sqrt{\frac{1}{S}}\frac{\frac{N_{z}}{W_{z}}\int_{z_{0}}^{z_{f}}dzJ_{\boldsymbol{k}}(z)e^{iK_{zd}z}}{\sqrt{\rho\int_{z_{0}}^{z_{f}}dz|J_{\boldsymbol{k}}(z)|^{2}}}.\label{eq:matrix_element-ref}
\end{align}
The quasi-momentum conservation $\delta_{\boldsymbol{K}_{\perp},\boldsymbol{k}}$
terms in Eqs. (\ref{eq:matrix_element-1}) and (\ref{eq:matrix_element-ref})
reveal that a plexciton with in-plane wavevector $\boldsymbol{k}$
will emit UHP photons with the same in-plane wavevector.

To begin our evaluation of the rates, we plug Eq. (\ref{eq:matrix_element-1})
into Eq. (\ref{eq:gamma_0}) and simplify the resulting expression,

\begin{eqnarray}
\gamma_{y_{\boldsymbol{k}},0} & = & \frac{4\pi}{\hbar}\sum_{\boldsymbol{K}_{\perp},\chi}\sum_{K_{zd}<0}\left[\left|\frac{\zeta_{y_{\boldsymbol{k}}}^{(exc)}}{\sqrt{1+|r_{\boldsymbol{K}\chi}|^{2}}}\sqrt{\frac{\hbar\omega_{\boldsymbol{K}}^{UHP}}{2\epsilon_{0}\epsilon_{d}V}}\boldsymbol{\mu}_{ge}\cdot\boldsymbol{e}_{\boldsymbol{K},\chi}^{*}N_{x}N_{y}\delta_{\boldsymbol{K}_{\perp},\boldsymbol{k}}\sqrt{\frac{1}{S}}\frac{\frac{N_{z}}{W_{z}}\int_{z_{0}}^{z_{f}}J_{\boldsymbol{k}}(z_{s})e^{-iK_{zd}z_{s}}}{\sqrt{\rho\int_{z_{0}}^{z_{f}}dz|J_{\boldsymbol{k}}(z)|^{2}}}\right|^{2}\right.\nonumber \\
 &  & \left.+\left|\frac{r_{\boldsymbol{K}\chi}^{*}\zeta_{y_{\boldsymbol{k}}}^{(exc)}}{\sqrt{1+|r_{\boldsymbol{K}\chi}|^{2}}}\sqrt{\frac{\hbar\omega_{\boldsymbol{K}}^{UHP}}{2\epsilon_{0}\epsilon_{d}V}}\boldsymbol{\mu}_{ge}\cdot\boldsymbol{e}_{\boldsymbol{K}_{\perp}-K_{zd}\hat{\boldsymbol{z}},\chi}^{*}N_{x}N_{y}\delta_{\boldsymbol{K}_{\perp},\boldsymbol{k}}\sqrt{\frac{1}{S}}\frac{\frac{N_{z}}{W_{z}}\int_{z_{0}}^{z_{f}}J_{\boldsymbol{k}}(z_{s})e^{iK_{zd}z_{s}}}{\sqrt{\rho\int_{z_{0}}^{z_{f}}dz|J_{\boldsymbol{k}}(z)|^{2}}}\right|^{2}\right]\delta(\hbar\omega_{y_{\boldsymbol{k}}}-\hbar\omega_{\boldsymbol{K}}^{UHP})\nonumber \\
 & = & \frac{|\zeta_{y_{\boldsymbol{k}}}^{(exc)}|^{2}}{(1+|r_{\boldsymbol{K}\chi}|^{2})}\frac{4\pi}{\hbar}\Bigg(\frac{L_{z}}{2\pi}\int_{-\infty}^{0}dK_{zd}\Bigg)\sum_{q=\pm}\sum_{\chi}\frac{\hbar\omega_{\boldsymbol{k},s}}{2\epsilon_{0}\epsilon_{d}V}|\boldsymbol{\mu}_{eg}\cdot\boldsymbol{e}_{\boldsymbol{K},\chi}|^{2}\nonumber \\
 &  & \times\frac{N_{x}^{2}N_{y}^{2}}{S}\Big(\frac{N_{z}}{W_{z}}\Big)^{2}\left[\frac{|\int_{z_{0}}^{z_{f}}dzJ_{\boldsymbol{k}}(z)e^{-iK_{zd}z}|^{2}+|r_{\boldsymbol{K}\chi}|^{2}|\int_{z_{0}}^{z_{f}}dzJ_{\boldsymbol{k}}(z)e^{iK_{zd}z}|^{2}}{\rho\int_{z_{0}}^{z_{f}}dz|J_{\boldsymbol{k}}(z)|^{2}}\right]\nonumber \\
 &  & \times\Theta\left(\omega_{\boldsymbol{k},y}-\frac{c}{\sqrt{\epsilon_{d}}}|\boldsymbol{k}|\right)\frac{\sqrt{|\boldsymbol{k}|^{2}+k_{dz}^{2}}\delta(K_{zd}+k_{dz})}{\frac{\hbar c}{\sqrt{\epsilon_{d}}}k_{dz}}\\
 & = & |\zeta_{y_{\boldsymbol{k}}}^{(exc)}|^{2}\left(\frac{\rho\omega_{\boldsymbol{k},y}^{2}}{2\epsilon_{0}\hbar c^{2}}\right)\Theta\left(\omega_{\boldsymbol{k},y}-\frac{c}{\sqrt{\epsilon_{d}}}|\boldsymbol{k}|\right)\frac{\sum_{q=\pm}\frac{1}{k_{dz}}\sum_{\chi}|\boldsymbol{\mu}_{eg}\cdot\boldsymbol{e}_{\boldsymbol{k}+qk_{dz}\hat{\boldsymbol{z}},\chi}|^{2}|\int_{z_{0}}^{z_{f}}dzJ_{\boldsymbol{k}}(z)e^{-iqk_{zd}z}|^{2}}{\int_{z_{0}}^{z_{f}}dz|J_{\boldsymbol{k}}(z)|^{2}}.\label{eq:gamma_0_final}
\end{eqnarray}
To go from the first to the second equality, we converted $\sum_{K_{zd}<0}\to\frac{L_{z}}{2\pi}\int_{-\infty}^{0}dK_{zd}$,
where $V=SL_{z}$ is the UHP quantization volume. Note that while
we considered the quantization area to be equal to the plexciton setup
surface area $S$, the quantization box vertical dimension $L_{z}$
is not equal to the thickness of the organic layer $W_{z}$ in general.
We have also used the $\delta_{\boldsymbol{K}_{\perp},\boldsymbol{k}}$
condition to rewrite the energy-conservation restriction as a constraint
on the vertical momentum of the emited photon $\delta(\hbar\omega_{\boldsymbol{k},y}-\hbar\omega_{\boldsymbol{K}}^{UHP})=\Theta\left(\omega_{\boldsymbol{k},s}-\frac{c}{\sqrt{\epsilon_{d}}}|\boldsymbol{k}|\right)\frac{\sqrt{|\boldsymbol{k}|^{2}+k_{dz}^{2}}[\delta(K_{zd}-k_{zd})+\delta(K_{zd}+k_{zd})]}{\frac{\hbar c}{\sqrt{\epsilon_{d}}}k_{dz}}$,
where we have used the shorthand notation
\begin{equation}
k_{zd}=\sqrt{\frac{\epsilon_{d}\omega_{\boldsymbol{k},y}^{2}}{c^{2}}-|\boldsymbol{k}|^{2}}>0.\label{eq:kzd}
\end{equation}
Finally, we used the fact that for a lossless metal ($\Gamma=0$)
and for $\omega_{\boldsymbol{K}}^{UHP}<\frac{\omega_{P}}{\sqrt{\epsilon_{d}+\epsilon_{\infty}}}$
(the UP frequency is below the asymptotic SP frequency for $k\to\infty$,
which is typically the case), $K_{zd}\in\Re$ and $K_{zm}\in\Im$,
implying that $|r_{\boldsymbol{K}\chi}|^{2}=1$,\emph{ i.e.}, metals
are perfectly reflective. Proceeding analogously by plugging Eqs.
(\ref{eq:matrix_element-1}) and (\ref{eq:matrix_element-ref}) into
Eq. (\ref{eq:gamma_ref}), we obtain,

\begin{subequations}\label{eq:gamma_final_exps}

\begin{eqnarray}
\gamma_{y_{\boldsymbol{k}},ref} & = & |\zeta_{y_{\boldsymbol{k}}}^{(exc)}|^{2}\left(\frac{\rho\omega_{\boldsymbol{k},y}^{2}}{2\epsilon_{0}\hbar c^{2}}\right)\Theta\left(\omega_{\boldsymbol{k},y}-\frac{c}{\sqrt{\epsilon_{d}}}|\boldsymbol{k}|\right)\nonumber \\
 &  & \times\frac{\Re\Bigg[2\sum_{\chi}\frac{1}{k_{dz}}r_{\boldsymbol{k}-k_{dz}\hat{\boldsymbol{z}},\chi}(\boldsymbol{e}_{\boldsymbol{k}+k_{dz}\hat{\boldsymbol{z}},\chi}\cdot\boldsymbol{\mu}_{eg})(\boldsymbol{\mu}_{ge}\cdot\boldsymbol{e}_{\boldsymbol{k}-k_{dz}\hat{\boldsymbol{z}},\chi}^{*})\int_{z_{0}}^{z_{f}}dzJ_{\boldsymbol{k}}^{*}(z)e^{ik_{zd}z}\int_{z_{0}}^{z_{f}}dz'J_{\boldsymbol{k}}(z')e^{ik_{zd}z'}\Bigg]}{\int_{z_{0}}^{z_{f}}dz|J_{\boldsymbol{k}}(z)|^{2}}.\label{eq:gamma_ref_final}
\end{eqnarray}

\end{subequations}

\subsection{Connection with Green's function formalism}

\subsubsection{Dyadic Green's function}

The quantum dyadic Green's function $\overleftrightarrow{G}$ is the
half-sided Fourier transform of the correlation function of the electric
field in the region of interest, $\hat{\boldsymbol{\mathcal{E}}}_{UHP}(\boldsymbol{r},t)=\sum_{\mu}\sqrt{\frac{\hbar\omega_{\mu}}{2\epsilon_{0}}}b_{\mu}(t)\boldsymbol{F}_{\mu}(\boldsymbol{r})+\text{h.c.}$
\cite{heitler1954quantum,sipe},

\begin{align}
G(\boldsymbol{r},\boldsymbol{r}',\omega) & =\frac{ic^{2}\epsilon_{0}}{\hbar\omega^{2}}\int_{0}^{\infty}dte^{i\omega t}\langle[\hat{\boldsymbol{\mathcal{E}}}_{UHP}(\boldsymbol{r},t),\hat{\boldsymbol{\mathcal{E}}}_{UHP}(\boldsymbol{r'},0)]\rangle\nonumber \\
 & =\text{lim}_{\epsilon\to0^{+}}\sum_{\eta}c^{2}\frac{\boldsymbol{F}_{\eta}(\boldsymbol{r})\otimes\boldsymbol{F}_{\eta}^{*}(\boldsymbol{r}')}{\omega_{\eta}^{2}-(\omega+i\epsilon)^{2}},\label{eq:dyadic_quantum}
\end{align}
where $\boldsymbol{F}_{\mu}(\boldsymbol{r})$ is the electric field
profile of the $\eta$th mode satisfying the orthonormality relation
$\int d^{3}r\epsilon(\boldsymbol{r})\boldsymbol{F}_{\eta}(\boldsymbol{r})\boldsymbol{F}_{\eta'}^{*}(\boldsymbol{r})=\delta_{\eta\eta'}$.
From Eq. (\ref{eq:dyadic_quantum}) it follows that \cite{novotny}

\begin{equation}
\Im G_{\beta\alpha}(\boldsymbol{r},\boldsymbol{r}',\omega)=\frac{\pi c^{2}}{2\omega}\sum_{\eta}[\boldsymbol{F}_{\eta}(\boldsymbol{r})]_{\beta}[\boldsymbol{F}_{\eta}^{*}(\boldsymbol{r}')]_{\alpha}\delta(\omega_{\mu}-\omega),\label{eq:imG}
\end{equation}
where we have ignored the negative frequency poles. Fermi's golden
rule rate in Eq. (\ref{eq:gamma_sk}) can be readily expressed in
terms of $G_{\beta\alpha}$ using Eq. (\ref{eq:imG}),

\begin{align}
\gamma_{y_{\boldsymbol{k}}}^{Green} & =\frac{2\pi}{\hbar}\sum_{\eta}|\langle\mbox{vac};\eta|-\sum_{\boldsymbol{n}s}\hat{\boldsymbol{\mu}}_{\boldsymbol{n}s}\cdot\hat{\boldsymbol{\mathcal{E}}}_{UHP}(\boldsymbol{r}_{\boldsymbol{n}s})|y_{\boldsymbol{k}};0_{UHP}\rangle|^{2}\delta(\hbar\omega_{\eta}-\hbar\omega_{y_{\boldsymbol{k}}})\nonumber \\
 & =\frac{2\pi}{\hbar}|\zeta_{y_{\boldsymbol{k}}}^{(exc)}|^{2}\sum_{\eta}\sum_{\boldsymbol{n}s;\boldsymbol{n'}s'}\frac{\hbar\omega_{\mu}}{2\epsilon_{0}}d_{\boldsymbol{n}s}^{(\boldsymbol{k})*}\sum_{\alpha,\beta}\mu_{eg}^{\beta}[\boldsymbol{F}_{\eta}(\boldsymbol{r}_{\boldsymbol{n}s})]_{\beta}[\boldsymbol{F}_{\eta}^{*}(\boldsymbol{r}_{\boldsymbol{n'}s'})]_{\alpha}\mu_{ge}^{\alpha}d_{\boldsymbol{n'}s'}^{(\boldsymbol{k})}\delta(\hbar\omega_{\eta}-\hbar\omega_{y_{\boldsymbol{k}}})\nonumber \\
 & =|\zeta_{y_{\boldsymbol{k}}}^{(exc)}|^{2}\left(\frac{2\omega_{\boldsymbol{k},y}^{2}}{\epsilon_{0}\hbar c^{2}}\right)\Im\sum_{\boldsymbol{n}s;\boldsymbol{n'}s'}d_{\boldsymbol{n}s}^{(\boldsymbol{k})*}\Bigg[\sum_{\alpha,\beta}\mu_{eg}^{\beta}G_{\beta\alpha}(\boldsymbol{r}_{\boldsymbol{n}s},\boldsymbol{r}_{\boldsymbol{n'}s'},\omega_{\boldsymbol{k},s})\mu_{ge}^{\alpha}\Bigg]d_{\boldsymbol{n'}s'}^{(\boldsymbol{k})}.\label{eq:gamma_green}
\end{align}
Owing to the harmonic nature of electromagnetic excitations, the quantum
dyadic Green's function in Eq. (\ref{eq:dyadic_quantum}) is equivalent
to its classical counterpart \cite{heitler1954quantum,sipe}. The
textbook expression for the classical dyadic corresponding to our
problem is \cite{kong1975theory,Tai1994,novotny},
\begin{equation}
\overleftrightarrow{G}_{text}=\overleftrightarrow{G}_{0}(\boldsymbol{r}_{\boldsymbol{n}s},\boldsymbol{r}_{\boldsymbol{n}'s'},\omega)+\overleftrightarrow{G}_{ref}(\boldsymbol{r}_{\boldsymbol{n}s},\boldsymbol{r}_{\boldsymbol{n}'s'},\omega).\label{eq:G=00003DG0+Gref}
\end{equation}
As we will next show, inserting Eq. (\ref{eq:G=00003DG0+Gref}) into
Eq. (\ref{eq:gamma_green}) gives rise to identical expressions to
Eq. (\ref{eq:gamma_0_final}) and (\ref{eq:gamma_ref_final}), representing
direct and interference terms involving free-space and reflected waves,
respectively.

$\overleftrightarrow{G}_{0}$ corresponds to the zeroth order free-space
propagation of the electromagnetic field in the absence of the metal
surface,

\begin{eqnarray}
\overleftrightarrow{G}_{0}(\boldsymbol{r}_{\boldsymbol{n}s},\boldsymbol{r}_{\boldsymbol{n}'s'},\omega) & = & \frac{i}{8\pi^{2}}\int_{-\infty}^{\infty}dK_{x}\int_{-\infty}^{\infty}dK_{y}e^{i[\boldsymbol{K}_{\perp}\cdot(\boldsymbol{r}_{\boldsymbol{n}}-\boldsymbol{r}_{\boldsymbol{n'}})+k_{zd}|z_{s}-z_{s'}|]}\overleftrightarrow{M}_{0}(\omega,\boldsymbol{K}_{\perp},z_{s},z_{s'}),\label{eq:G0}
\end{eqnarray}
where $\overleftrightarrow{M}_{0}(\omega,\boldsymbol{K}_{\perp},z_{s},z_{s'})=\overleftrightarrow{M}_{0\pm}(\omega,\boldsymbol{K}_{\perp})=\overleftrightarrow{M}_{0\pm}^{s}(\omega,\boldsymbol{K}_{\perp})+\overleftrightarrow{M}_{0\pm}^{p}(\omega,\boldsymbol{K}_{\perp})$.
Here, the positive sign is chosen if $z_{s}>z_{s'}$ and the negative
one if $z_{s}<z_{s'}$; the average $\overleftrightarrow{M}_{0}(\omega,\boldsymbol{K}_{\perp},z_{s},z_{s'})=\frac{1}{2}\Big[\overleftrightarrow{M}_{0+}(\omega,\boldsymbol{K}_{\perp})+\overleftrightarrow{M}_{0-}(\omega,\boldsymbol{K}_{\perp})\Big]$
is evaluated if $z_{s}=z_{s'}$. These matrices can be written as
outer products of the free-space photon polarizations, $\overleftrightarrow{M}_{0q,\beta\alpha}^{\chi}(\omega,\boldsymbol{K}_{\perp})=\frac{1}{k_{zd}}(\boldsymbol{e}_{\boldsymbol{K}_{\perp}+qk_{zd}\hat{\boldsymbol{z}},\chi})_{\beta}(\boldsymbol{e}_{\boldsymbol{K}_{\perp}+qk_{zd}\hat{\boldsymbol{z}},\chi})_{\alpha}^{*}$
\cite{kong1975theory}. More explicitly \cite{novotny},

\begin{equation}
\overleftrightarrow{M}_{0\pm}=\frac{1}{K_{d}^{2}k_{zd}}\left[\begin{array}{ccc}
K_{d}^{2}-K_{x}^{2} & -K_{x}K_{y} & \mp K_{x}k_{zd}\\
-K_{x}K_{y} & K_{d}^{2}-K_{y}^{2} & \mp K_{y}k_{zd}\\
\mp K_{x}k_{zd} & \mp K_{y}k_{zd} & K_{d}^{2}-k_{zd}^{2}
\end{array}\right].\label{eq:M0pm}
\end{equation}
We have used the notation $K_{d}(\omega)=|\boldsymbol{K}(\omega)|=\frac{\sqrt{\epsilon_{d}}\omega}{c}$.
On the other hand, $\overleftrightarrow{G}_{ref}$ corresponds to
the reflected electromagnetic waves from the metal-dielectric interface,

\begin{equation}
\overleftrightarrow{G}_{ref}(\boldsymbol{r}_{\boldsymbol{n}s},\boldsymbol{r}_{\boldsymbol{n}'s'},\omega)=\frac{i}{8\pi^{2}}\int_{|\boldsymbol{K}|<\frac{\sqrt{\epsilon_{d}}\omega}{c}}d^{2}\boldsymbol{K}e^{i[\boldsymbol{K}\cdot(\boldsymbol{r}_{\boldsymbol{n}}-\boldsymbol{r}_{\boldsymbol{n'}})+k_{zd}(z_{s}+z_{s'})]}\overleftrightarrow{M}_{ref}(\omega,\boldsymbol{k}),\label{eq:Gref}
\end{equation}
with $\overleftrightarrow{M}_{ref}(\omega,\boldsymbol{k})=\overleftrightarrow{M}_{ref}^{s}(\omega,\boldsymbol{k})+\overleftrightarrow{M}_{ref}^{p}(\omega,\boldsymbol{k})$,
where $\overleftrightarrow{M}_{ref,\beta\alpha}^{\chi}(\omega,\boldsymbol{K}_{\perp})=\frac{r_{\boldsymbol{K},\chi}}{K_{zd}}(\boldsymbol{e}_{\boldsymbol{K}_{\perp}+k_{zd}\hat{\boldsymbol{z}},\chi})_{\beta}(\boldsymbol{e}_{\boldsymbol{K}_{\perp}-k_{zd}\hat{\boldsymbol{z}},\chi})_{\alpha}^{*}$
\cite{kong1975theory}, where \cite{novotny}

\begin{eqnarray}
\overleftrightarrow{M}_{ref}^{s}(\omega,\boldsymbol{K}) & = & \frac{r_{\boldsymbol{K},s}}{k_{zd}(K_{x}^{2}+K_{y}^{2})}\left[\begin{array}{ccc}
K_{y}^{2} & -K_{x}K_{y} & 0\\
-K_{x}K_{y} & K_{x}^{2} & 0\\
0 & 0 & 0
\end{array}\right],\label{eq:Mref_s}\\
\overleftrightarrow{M}_{ref}^{p}(\omega,\boldsymbol{k}) & = & \frac{-r_{\boldsymbol{K},p}}{K_{d}^{2}(K_{x}^{2}+K_{y}^{2})}\left[\begin{array}{ccc}
K_{x}^{2}k_{zd} & K_{x}K_{y}k_{zd} & K_{x}(K_{x}^{2}+K_{y}^{2})\\
K_{x}K_{y}k_{zd} & K_{y}^{2}k_{zd} & K_{y}(K_{x}^{2}+K_{y}^{2})\\
-K_{x}(K_{x}^{2}+K_{y}^{2}) & -K_{y}(K_{x}^{2}+K_{y}^{2}) & -(K_{x}^{2}+K_{y}^{2})^{2}/k_{zd}
\end{array}\right],\label{eq:Mref_p}
\end{eqnarray}
Notice that we have limited Eq. (\ref{eq:Gref}) to the integration
$\int_{|\boldsymbol{K}|<\frac{\sqrt{\epsilon_{d}}\omega}{c}}d^{2}\boldsymbol{K}$
over radiative modes. For real valued relative permittivities, $\epsilon_{d}$
and $\epsilon_{m}$, integration over evanescent wavevectors (with
$K_{dz}\in\Im$) picks up the corresponding SP mode pole contributions.
We need not take these into account, since we have already accounted
for SP modes in our strong-coupling Hamiltonian formalism.

\subsubsection{Evaluation of Eq. (\ref{eq:gamma_green})}

We now proceed to directly evaluate Eq. (\ref{eq:gamma_green}) for
the different terms of $\overleftrightarrow{G}_{text}$ in Eq. (\ref{eq:G=00003DG0+Gref}).
Starting with the free-space term,

\begin{eqnarray}
 &  & \gamma_{y_{\boldsymbol{k}},0}^{Green}\nonumber \\
 & = & |\zeta_{y_{\boldsymbol{k}}}^{(exc)}|^{2}\left(\frac{2\omega_{\boldsymbol{k},y}^{2}}{\epsilon_{0}\hbar c^{2}}\right)\Im\sum_{\boldsymbol{n}s;\boldsymbol{n'}s'}d_{\boldsymbol{n}s}^{(\boldsymbol{k})*}d_{\boldsymbol{n'}s'}^{(\boldsymbol{k})}\left[\sum_{\alpha,\beta}\mu_{ge}^{\beta}G_{0,\beta\alpha}(\boldsymbol{r}_{\boldsymbol{n}s},\boldsymbol{r}_{\boldsymbol{n'}s'},\omega_{\boldsymbol{k},s})\mu_{eg}^{\alpha}\right]\nonumber \\
 & = & |\zeta_{y_{\boldsymbol{k}}}^{(exc)}|^{2}\left(\frac{2\omega_{\boldsymbol{k},y}^{2}}{\epsilon_{0}\hbar c^{2}}\right)\Im\sum_{\boldsymbol{n}s;\boldsymbol{n'}s'}\frac{1}{S}\frac{J_{\boldsymbol{k}}^{*}(z_{s})e^{-i\boldsymbol{k}\cdot\boldsymbol{r}_{\boldsymbol{n}}}J_{\boldsymbol{k}}(z_{s'})e^{i\boldsymbol{k}\cdot\boldsymbol{r}_{\boldsymbol{n'}}}}{\mathcal{J}_{\boldsymbol{k}}^{2}}\nonumber \\
 &  & \times\Bigg\{\frac{i}{8\pi^{2}}\int_{-\infty}^{\infty}dK_{x}\int_{-\infty}^{\infty}dK_{y}e^{i[\boldsymbol{K}\cdot(\boldsymbol{r}_{\boldsymbol{n}}-\boldsymbol{r}_{\boldsymbol{n'}})+K_{dz}|z_{s}-z_{s'}|]}\sum_{\alpha,\beta}\mu_{ge}^{\beta}M_{0,\beta\alpha}(\omega_{\boldsymbol{k},s},\tilde{\boldsymbol{k}},z_{s},z_{s'})\mu_{eg}^{\alpha}\Bigg\}\nonumber \\
 & = & |\zeta_{y_{\boldsymbol{k}}}^{(exc)}|^{2}\left(\frac{2\omega_{\boldsymbol{k},y}^{2}}{\epsilon_{0}\hbar c^{2}}\right)\Im\sum_{ss'}\frac{1}{S}\frac{J_{\boldsymbol{k}}^{*}(z_{s})J_{\boldsymbol{k}}(z_{s'})}{\mathcal{J}_{\boldsymbol{k}}^{2}}\Bigg[\frac{i}{8\pi^{2}}\int_{-\infty}^{\infty}dK_{x}\int_{-\infty}^{\infty}dK_{y}N_{x}^{2}N_{y}^{2}\delta_{\boldsymbol{k},\boldsymbol{K}}e^{iK_{dz}|z_{s}-z_{s'}|}\sum_{\alpha,\beta}\mu_{ge}^{\beta}M_{0,\beta\alpha}(\omega_{\boldsymbol{k},s},\tilde{\boldsymbol{k}},z_{s},z_{s'})\mu_{eg}^{\alpha}\Bigg]\nonumber \\
 & = & |\zeta_{y_{\boldsymbol{k}}}^{(exc)}|^{2}\left(\frac{2\omega_{\boldsymbol{k},y}^{2}}{\epsilon_{0}\hbar c^{2}}\right)\Im\sum_{ss'}\frac{1}{S}\frac{J_{\boldsymbol{k}}^{*}(z_{s})J_{\boldsymbol{k}}(z_{s'})N_{x}^{2}N_{y}^{2}}{\mathcal{J}_{\boldsymbol{k}}^{2}}\nonumber \\
 &  & \times\Bigg[\frac{i}{8\pi^{2}}\int_{-\infty}^{\infty}dK_{x}\int_{-\infty}^{\infty}dK_{y}\frac{(2\pi)^{2}}{S}\delta(\boldsymbol{k}-\boldsymbol{K})e^{ik_{dz}|z_{s}-z_{s'}|}\sum_{\alpha,\beta}\mu_{ge}^{\beta}M_{0,\beta\alpha}(\omega_{\boldsymbol{k},s},\tilde{\boldsymbol{k}},z_{s},z_{s'})\mu_{eg}^{\alpha}\Bigg]\nonumber \\
 & = & |\zeta_{y_{\boldsymbol{k}}}^{(exc)}|^{2}\left(\frac{\omega_{\boldsymbol{k},y}^{2}}{2\epsilon_{0}\hbar c^{2}}\right)\frac{2\Re\Bigg\{\Big[\sum_{\alpha,\beta}\mu_{ge}^{\beta}M_{0,\beta\alpha}(\omega_{\boldsymbol{k},s},\boldsymbol{k},z_{s},z_{s'})\mu_{eg}^{\alpha}\Big]\sum_{ss'}J_{\boldsymbol{k}}^{*}(z_{s})J_{\boldsymbol{k}}(z_{s'})e^{ik_{dz}|z_{s}-z_{s'}|}\Bigg\}}{\mathcal{J}_{\boldsymbol{k}}^{2}}\Bigg(\frac{N_{x}^{2}N_{y}^{2}}{S^{2}}\Bigg).\label{eq:gamma_sk0_first_steps}
\end{eqnarray}
Going from the third to the fourth line, we have approximated $\delta_{\boldsymbol{k},\boldsymbol{K}}=\frac{(2\pi)^{2}}{S}\delta(\boldsymbol{k}-\boldsymbol{K})$.
At a first glance, it is not clear that Eq. (\ref{eq:gamma_sk0_first_steps})
is equal to Eq. (\ref{eq:gamma_0_final}). However, with some algebraic
effort, it is possible to establish the following identity,
\begin{eqnarray}
 &  & 2\Re\Bigg[\Bigg(\sum_{\alpha,\beta}\mu_{ge}^{\beta}M_{0,\beta\alpha}(\omega_{\boldsymbol{k},y},\boldsymbol{k},z_{s},z_{s'})\mu_{eg}^{\alpha}\Bigg)\sum_{s,s'}J_{\boldsymbol{k}}^{*}(z_{s})J_{\boldsymbol{k}}(z_{s'})e^{ik_{dz}|z_{s}-z_{s'}|}\Bigg]\nonumber \\
 & = & \Theta\left(\omega_{\boldsymbol{k},y}-\frac{c}{\sqrt{\epsilon_{d}}}|\boldsymbol{k}|\right)\sum_{q=\pm}\Bigg[\sum_{\alpha,\beta}\mu_{ge}^{\beta}M_{0q,\beta\alpha}(\omega_{\boldsymbol{k},s},\boldsymbol{k})\mu_{eg}^{\alpha}|\sum_{s}J_{\boldsymbol{k}}(z_{s})e^{-iqk_{dz}z}|^{2}\Bigg]\nonumber \\
 & = & \Theta\left(\omega_{\boldsymbol{k},y}-\frac{c}{\sqrt{\epsilon_{d}}}|\boldsymbol{k}|\right)\sum_{q=\pm}\Bigg[\frac{1}{k_{dz}}\sum_{\chi}|\boldsymbol{\mu}_{eg}\cdot\boldsymbol{e}_{\boldsymbol{k}+qk_{dz}\hat{\boldsymbol{z}},\chi}|^{2}|\sum_{s}J_{\boldsymbol{k}}(z_{s})e^{-iqk_{dz}z}|^{2}\Bigg],\label{eq:identity_dyadic}
\end{eqnarray}
where we have used the definition of $M_{0q,\beta\alpha}(\omega_{\boldsymbol{k},y},\boldsymbol{k})$
presented above. Substituting Eq. (\ref{eq:identity_dyadic}) together
with $\sum_{s}\to\frac{N_{z}}{W_{z}}\int_{z_{0}}^{z_{f}}dz$ on Eq.
(\ref{eq:gamma_sk0_first_steps}) yields $\gamma_{y_{\boldsymbol{k}},0}^{Green}=\gamma_{y_{\boldsymbol{k}},0}$,
thus verifying Eq. (\ref{eq:gamma_0_final}). Interestingly, the physical
interpretations of these two equivalent calculations is slightly different:
$\gamma_{y_{\boldsymbol{k}},0}$ is the sum of incoherent contributions
from free and reflected waves (see Eq. (\ref{eq:gamma_0})) while
$\gamma_{y_{\boldsymbol{k}},0}^{Green}$ uses information of free
waves alone (see Eqs. (\ref{eq:G0}) and (\ref{eq:M0pm})). These
two physical pictures can be reconciled by noting the form of the
modal decomposition of $\hat{\boldsymbol{\mathcal{E}}}_{UHP}$ in
Eq. (\ref{eq:E_UHP}). Using similar manipulations, it follows straightforwardly
that $\gamma_{y_{\boldsymbol{k}},ref}^{Green}=\gamma_{y_{\boldsymbol{k}},ref}$
as in Eq. (\ref{eq:gamma_ref_final}). We thus obtain $\gamma_{y_{\boldsymbol{k}}}^{Green}=\gamma_{y_{\boldsymbol{k}}}.$

\subsection{Final expressions}

We aim to provide simplified expressions that specialize to a common
experimental situation where molecules are randomly oriented in the
organic layer. Denoting isotropic averages $\langle\cdot\rangle_{iso}$,
the following identities will be useful,

\begin{subequations}\label{eq:iso}

\begin{align}
\left\langle \sum_{q=\pm}\frac{1}{k_{dz}}\sum_{\chi}|\boldsymbol{\mu}_{eg}\cdot\boldsymbol{e}_{\boldsymbol{k}+qk_{dz}\hat{\boldsymbol{z}},\chi}|^{2}\right\rangle _{iso} & =|\boldsymbol{\mu}_{eg}|^{2}\sum_{q=\pm}\text{Tr}\Big[M_{0q,\beta\alpha}(\omega_{\boldsymbol{k},s},\boldsymbol{k})\Big]\nonumber \\
 & =\frac{4|\boldsymbol{\mu}_{eg}|^{2}}{3k_{zd}},\label{eq:iso_0}\\
\left\langle \frac{2}{k_{dz}}\sum_{\chi}r_{\boldsymbol{k}-k_{dz}\hat{\boldsymbol{z}},\chi}(\boldsymbol{e}_{\boldsymbol{k}+k_{dz}\hat{\boldsymbol{z}},\chi}\cdot\boldsymbol{\mu}_{eg})(\boldsymbol{\mu}_{ge}\cdot\boldsymbol{e}_{\boldsymbol{k}-k_{dz}\hat{\boldsymbol{z}},\chi}^{*})\right\rangle _{iso} & =2|\boldsymbol{\mu}_{eg}|^{2}\text{Tr}\Big[M_{ref,\beta\alpha}^{\chi}(\omega_{\boldsymbol{k},s},\boldsymbol{k})\Big]\nonumber \\
 & =\frac{2\left(r_{\boldsymbol{k}-k_{dz}\hat{\boldsymbol{z}},s}-r_{\boldsymbol{k}-k_{dz}\hat{\boldsymbol{z}},p}\frac{k_{dz}^{2}-|\boldsymbol{k}|^{2}}{k_{dz}^{2}+|\boldsymbol{k}|^{2}}\right)|\boldsymbol{\mu}_{eg}|^{2}}{3k_{zd}}.\label{eq:iso_cross}
\end{align}

\end{subequations}\noindent Plugging these expressions into Eqs.
(\ref{eq:gamma_0_final})\textendash (\ref{eq:gamma_ref}), we obtain
Eqs. (\ref{eq:ratio})\textendash (\ref{eq:f}). The evaluation of
the associated integrals is analytically tractable,

\begin{subequations}\label{eq:integrals_J}

\begin{eqnarray}
\int_{z_{0}}^{z_{f}}dzJ_{\boldsymbol{k}}(z)e^{-iqk_{dz}z} & = & \int_{z_{0}}^{z_{f}}dz\boldsymbol{\mu}_{eg}\cdot\boldsymbol{E}_{\boldsymbol{k}}\sqrt{\frac{\hbar\omega_{\boldsymbol{k}}^{SP}}{2\epsilon_{0}L_{\boldsymbol{k}}}}e^{-(\alpha_{\boldsymbol{k}d}+iqk_{dz})z}\nonumber \\
 & = & \boldsymbol{\mu}_{eg}\cdot\boldsymbol{E}_{\boldsymbol{k}}\sqrt{\frac{\hbar\omega_{\boldsymbol{k}}^{SP}}{2\epsilon_{0}L_{\boldsymbol{k}}}}\Bigg(\frac{e^{-(\alpha_{\boldsymbol{k}d}+iqk_{z1})z_{0}}-e^{-(\alpha_{\boldsymbol{k}d}+iqk_{z1})z_{f}}}{\alpha_{\boldsymbol{k}d}+iqk_{dz}}\Bigg),\label{eq:integral1}\\
\int_{z_{0}}^{z_{f}}dz|J_{\boldsymbol{k}}(z)|^{2} & = & \int_{z_{0}}^{z_{f}}dz\left|\boldsymbol{\mu}_{eg}\cdot\boldsymbol{E}_{\boldsymbol{k}}\sqrt{\frac{\hbar\omega_{\boldsymbol{k}}^{SP}}{2\epsilon_{0}L_{\boldsymbol{k}}}}e^{-\alpha_{\boldsymbol{k}d}z}\right|^{2}\nonumber \\
 & = & |\boldsymbol{\mu}_{eg}\cdot\boldsymbol{E}_{\boldsymbol{k}}|^{2}\frac{\hbar\omega_{\boldsymbol{k}}^{SP}}{2\epsilon_{0}L_{\boldsymbol{k}}}\Bigg(\frac{e^{-2\alpha_{\boldsymbol{k}d}z_{0}}-e^{-2\alpha_{\boldsymbol{k}d}z_{f}}}{2\alpha_{\boldsymbol{k}d}}\Bigg).\label{eq:integral2}
\end{eqnarray}

\end{subequations}\noindent where the terms $\boldsymbol{\mu}_{eg}\cdot\boldsymbol{E}_{\boldsymbol{k}}\sqrt{\frac{\hbar\omega_{\boldsymbol{k}}^{SP}}{2\epsilon_{0}L_{\boldsymbol{k}}}}$
in these expressions are unimportant in Eq. (\ref{eq:ratio}) since
they cancel out.

\section{Effects of metal losses \label{sec:Effects-of-metal}}

In the main text, we have established a condition upon which we expect
the superradiant PL rates to be observed experimentally, namely, that
the critical wavevector $\boldsymbol{k}^{*}$ is close enough to the
anticrossing so the associated plexciton state can be actually considered
to arise from strong SP-exciton coupling and the presented polariton
theory holds true. In this Section, we consider the robustness of
these PL rates against metal losses provided the latter condition
is fullfilled.

\begin{figure}
\centering{}\includegraphics[scale=0.3]{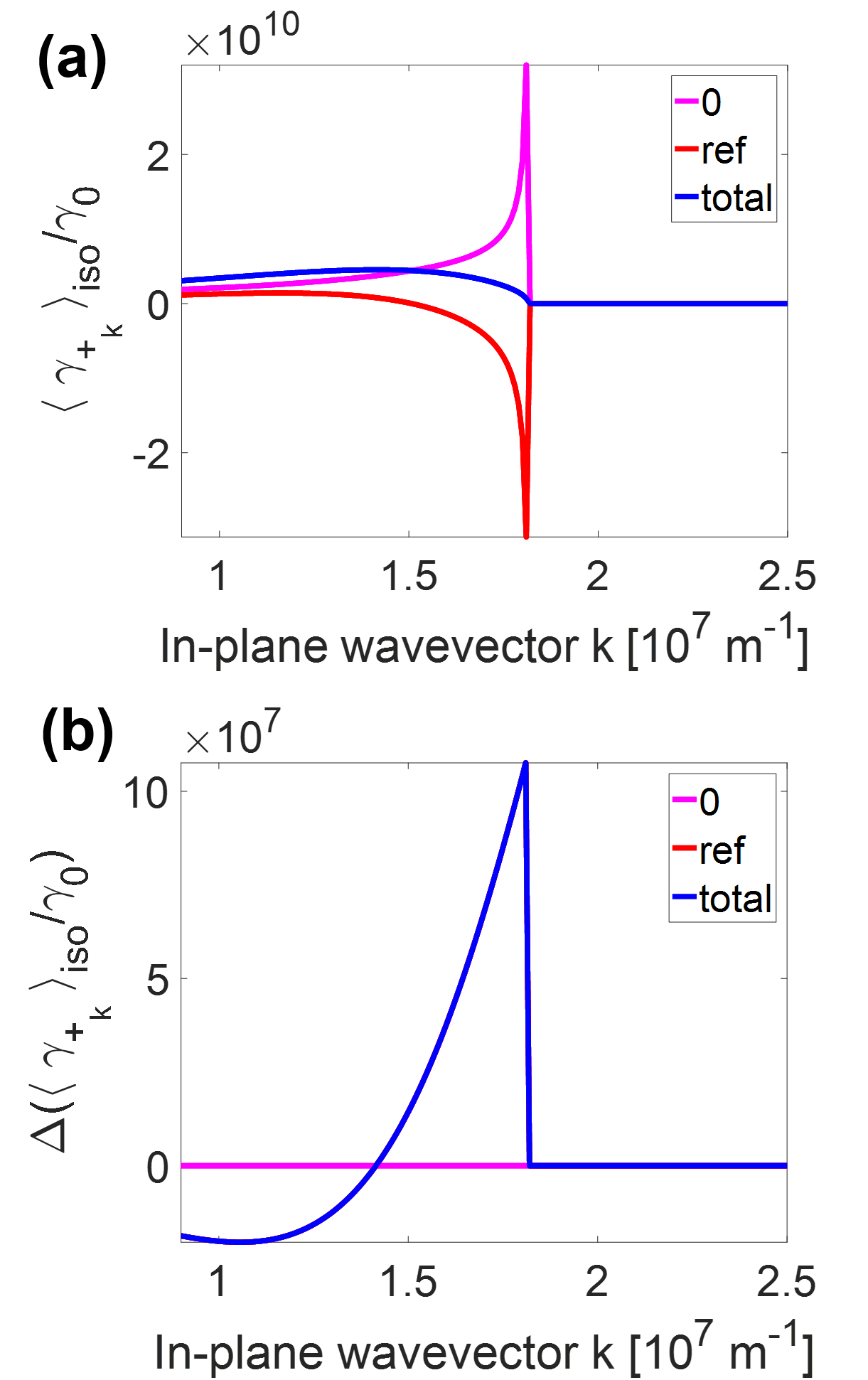}\caption{\emph{Effects of metal dissipation in plexciton PL rates.} (a) PL
rates for same conditions as in Fig. \ref{fig:dispersion}b, but with
lossy metal. No qualitative differences appear when compared to lossless
metal case; importantly, van Hove anomalies for $\langle\gamma_{+_{\boldsymbol{k}^{*}},i}\rangle_{iso}$
 remain because $\frac{dk_{zd}}{d\omega}|_{\omega=\omega_{y_{\boldsymbol{k^{*}}}}}$
still diverges. (b) A close-up on the quantitative PL rate differences
$\Delta\frac{\langle\gamma_{y_{\boldsymbol{k}},i}\rangle_{iso}}{\gamma_{SM,0}}$
with and without metal loss; these differences are negligible when
compared to the scale in Fig. \ref{fig:dispersion}b or in (a). Note
that $\Delta\frac{\langle\gamma_{y_{\boldsymbol{k}},0}\rangle_{iso}}{\gamma_{SM,0}}=0$
because it does not depend on the Fresnel coefficient $r_{\boldsymbol{\boldsymbol{k}-}k_{zd}\hat{\boldsymbol{z}},\chi}$,
which contains information about metal loss, so $\Delta\frac{\langle\gamma_{y_{\boldsymbol{k}}}\rangle_{iso}}{\gamma_{SM,0}}=\Delta\frac{\langle\gamma_{y_{\boldsymbol{k}},ref}\rangle_{iso}}{\gamma_{SM,0}}$.
\label{fig:sm}}
\end{figure}

A complex-valued metal permittivity $\epsilon_{m}(\omega)=\epsilon_{\infty}-\frac{\omega_{P}^{2}}{\omega^{2}+i\omega\gamma}$
for $\gamma\in\Re$, $\gamma>0$ modifies the modal decomposition
of $\hat{\boldsymbol{\mathcal{E}}}_{UHP}$ in Eq. (\ref{eq:E_UHP}).
To handle this difficulty, it is convenient to invoke the M-QED formalism
due to Dung, Knoll, and Welsch \cite{dung2000spontaneous} (see also
\cite{vogel2006quantum,scheel2008macroscopic}), which bypasses a
description of the electromagnetic field in terms of delocalized modes
and instead regards it as a polarizable lossy continuum where Eqs.
(\ref{eq:imG}) remains valid. Assuming that the plexciton eigenstates
of $H_{\boldsymbol{k}}$ (Eq. (\ref{eq:Hk}), still taking $\gamma=0$)
are good zeroth-order initial states in a Fermi's golden rule PL calculation,
it follows that the rate expression in Eq. (\ref{eq:gamma_green})
still holds, and thus, Eqs. (\ref{eq:ratio}) and (\ref{eq:f}) remain
valid. This is a good assumption given that in the strong-coupling
regime, the polariton energies are expected to be separated by a Rabi
splitting which is much larger than the SP linewidth, $\omega_{+_{\boldsymbol{k}}}-\omega_{-_{\boldsymbol{k}}}\gg\gamma$
\cite{torma2015strong,ribeiro2018polariton}. The only difference
with the previous calculations is that we must account for the fact
that metals are not perfect reflectors in the presence of loss, $|r_{\boldsymbol{\boldsymbol{k}-}k_{zd}\hat{\boldsymbol{z}},\chi}|^{2}\neq1$,
since $k_{zd}$ becomes complex valued (we need to make sure that
$\Im k_{zd}<0$ to guarantee that the modes vanish at $z\to-\infty$).
Fig. \ref{fig:sm}a shows the same plots of Fig. \ref{fig:dispersion}b
with the exception that $\gamma=0.07\,\text{eV}$ \cite{drude_parameters};
these figures are qualitatively indistinguishable. Fig. \ref{fig:sm}b
shows the quantitative difference $\Delta\frac{\langle\gamma_{y_{\boldsymbol{k}},i}\rangle_{iso}}{\gamma_{SM,0}}=\frac{\langle\gamma_{y_{\boldsymbol{k}},i}\rangle_{iso}}{\gamma_{SM,0}}\Big|_{\gamma=0.07\,\text{eV}}-\frac{\langle\gamma_{y_{\boldsymbol{k}},i}\rangle_{iso}}{\gamma_{SM,0}}\Big|_{\gamma=0}$,
which is negligible in the scale of each of the normalized PL rates.
It is interesting that the van Hove anomalies for $\langle\gamma_{y_{\boldsymbol{k}},i}\rangle_{iso}$
remain even for $\gamma\neq0$, given that they arise from purely
geometric (rather than dynamic) considerations. Thus, we conclude
that the predicted PL trends should be robust under a wide range of
experimental conditions.

\end{document}